\documentclass[review,3p,times]{elsarticle}
\usepackage[font=normalsize,labelfont=bf]{caption}
\usepackage{hyperref}
\journal{Computers \& Fluids}
\biboptions{sort&compress} 
\usepackage{xcolor}
\usepackage{amssymb,amsmath}
\usepackage{bm}

\usepackage{textcomp}
\usepackage{gensymb}
\usepackage{subcaption}
\usepackage{longtable,booktabs,tabularx,tabulary}
\newcolumntype{L}[1]{>{\hsize=#1\hsize\raggedright\arraybackslash}X}
\newcolumntype{R}[1]{>{\hsize=#1\hsize\raggedleft\arraybackslash}X}
\newcolumntype{C}[2]{>{\hsize=#1\hsize\columncolor{#2}\centering\arraybackslash}X}
\newcolumntype{Y}{>{\centering\arraybackslash}X}
\usepackage[status=draft,index,multiuser]{fixme}
\fxsetface{margin}{\linespread{1}\scriptsize}
\fxusetheme{color}

\FXRegisterAuthor{bz}{abz}{BZ}  
\FXRegisterAuthor{aw}{aaw}{AW}  
\FXRegisterAuthor{km}{akm}{KM}  
\FXRegisterAuthor{gb}{agb}{GB}  

\usepackage{soul}
\begin{document}
\begin{frontmatter}

\title{A voxelized immersed boundary (VIB) finite element method for accurate and efficient blood flow simulation}

\address[up]{University of Patras, Department of Agriculture, New buildings, 30200, Messolonghi, Greece}

\address[uwa]{Intelligent Systems for Medicine Laboratory,
The University of Western Australia,35 Stirling Highway, Perth, Western Australia}

\address[HU]{Laboratory of Applied Mathematics, School of Science and Technology, Hellenic Open University}

\address[phpat]{Department of Physics, University of Patras, Rion, GR 26504, Greece}

\address[patrascv]{Laboratory of Hydraulic Engineering, Department of Civil Engineering, University of Patras, 26500 Patras, Greece}

\address[ITE]{Institute of Chemical Engineering Sciences (ICE-HT), Foundation for Research and Technology, Hellas (FORTH), GR-26504 Patras, Greece}

\address[harvard]{Harvard Medical School, Boston, Massachusetts,
USA}

\author[up,uwa]{G. C. Bourantas\corref{mycorrespondingauthor}}
\ead{gbourantas@upatras.gr}

\author[uwa]{B. F. Zwick}

\author[HU]{D. S. Lampropoulos}

\author[phpat]{V. C. Loukopoulos}

\author[phpat]{K. Katsanos}

\author[patrascv]{A. A. Dimas}

\author[ITE]{V. N. Burganos}

\author[uwa]{A. Wittek}

\author[uwa,harvard]{K. Miller}

\cortext[mycorrespondingauthor]{Corresponding author}

\begin{abstract}
We present an efficient and accurate immersed
boundary (IB) finite element (FE) method for internal flow problems with complex geometries (e.g., blood flow in the vascular system). In this study, we use a voxelized flow domain (discretized with hexahedral and tetrahedral elements) instead of a box domain, which is frequently used in IB methods. The proposed method utilizes the well-established incremental pressure correction scheme (IPCS) FE solver, and the boundary condition-enforced IB (BCE-IB) method to numerically solve the transient, incompressible Navier--Stokes flow equations. We verify the accuracy of our numerical method using the analytical solution for the Poiseuille flow in a cylinder, and the available experimental data (laser Doppler velocimetry) for the flow in a three-dimensional 90° angle tube bend. We further examine the accuracy and applicability of the proposed method by considering flow within complex geometries, such as blood flow in aneurysmal vessels and the aorta, flow configurations that would otherwise be difficult to solve by most IB methods. Our method offers high accuracy, as demonstrated by the verification examples, and high applicability, as demonstrated through the solution of blood flow within complex geometry. The proposed method is efficient, since it is as fast as the traditional finite element method used to solve the Navier--Stokes flow equations, with a small overhead (not more than 5$\%$) due to the numerical solution of a linear system formulated for the IB method.
\end{abstract}


\end{frontmatter}



\section{Introduction}

Computational fluid dynamics (CFD) simulations are becoming a reliable tool for analysis of complex flow regimes in healthy and pathological blood vessels. To accurately model such flows, often a scale-resolving turbulence model is required, and high-order methods are used for discretizing and numerically solving the governing flow equations. However, high-order numerical methods often necessitates the use of structured grids, which may not be able to handle a variety of complex geometries that arise in arterial flow domains. Additionally, even for laminar flow regimes, where the flow regime is less complex, computing derived quantities such as wall shear stress (WSS), time average wall shear stress (TAWSS) and oscillatory shear index (OSI), requires a high-quality mesh, which is extremely difficult to obtain.

The immersed boundary (IB) methods emerged as an attractive approach because of its ability to efficiently handle complex moving and rotating geometries on structured Cartesian grids, and to simulate flow past rigid and moving/deforming bodies with irregular/complex shapes \citep{PESKIN1972252, lai_peskin_2000_immersed}. IB methods, in contrast to mesh-based methods, avoid tedious mesh generation, since they rely on Cartesian grids to solve the governing flow equations, and on a discrete set of points (often referred to as Lagrangian points) for the imposition of prescribed velocity boundary conditions. The IB methods are particularly efficient for moving objects and deforming geometries—they are attractive with moving boundaries as they avoid re-meshing—where it does not require the generation of a new mesh at each time step, but only the updated position of the points describing the immersed boundary. 

Immersed boundary method-based simulations are frequently used in external aerodynamics problems \citep{CHOI2007757, COLONIUS20082131,doi:10.1146/annurev.fluid.37.061903.175743}. However, their application to internal fluid flow in complex geometries, such as blood flow in arteries, remains challenging. Yokoi et al.\citep{YOKOI20051} simulated blood flow in a cerebral artery with multiple aneurysms using an immersed boundary method (they discretized the flow domain using a Cartesian grid). Although Yokoi et al.\ didn't mention the percentage of nodes outside the flow domain over the total number of grid nodes (they used 0.6 million Cartesian grid nodes in total), it was apparent that a large portion of grid nodes were in the exterior of the fluid domain. Delorme et al.\ \citep{DELORME2013408} performed Large Eddy Simulation (LES) studies of powered Fontan hemodynamics using a relatively short vena cava and long pulmonary arteries to reduce the percentage of Cartesian grid nodes located in the exterior of the flow domain. Despite that, the authors reported that a significant number of grid nodes were located outside of the flow domain. To extend IB methods to simulate complex arterial geometries, de Zélicourt et al.\ \citep{ZELICOURT20091749} simulated blood flow in a geometry of the total cavo-pulmonary connection (TCPC) using an unstructured Cartesian grid. However, studies that apply the IB methods to complex internal flow configurations remain scarce. The key reason for this is increase in the internal memory and computational efforts for single block (structured and unstructured) Cartesian grids that are often used in the Ib methods \citep{ZELICOURT20091749}. Additionally, handling complex geometries on a structured grid, all the inflow and outflow boundaries of the geometry have to terminate on the boundary faces of the global bounding box that encloses both the fluid and solid regions. Even if this requirement could be met in certain cases by truncating the complex geometry, in many cases it is not doable. To overcome the aforementioned problems and extend the applicability of the IB method to simulate blood flow in complex anatomies, authors in \citep{anupindi_etal_2013_novel} proposed a method that combined multi-block structured grid and IB method on an inherently parallel framework. The method enabled the simulation of fluid flow in complex geometries, reducing the percentage of grid nodes located outside the immersed boundary.

In recent decades, a large class of IB methods based on the finite element (FE) method has been developed. Löhner et al. \citep{Lohner1, Lohner2} applied the kinetic and kinematic impositions of boundary conditions introduced in immersed boundary methods \citep{SOTIR} to adaptive nodal finite element grids. Glowinski et al. \cite{GLOW1,GLOW2} simulated viscous flow interacting with rigid particles by using a distributed Lagrange multiplier field to enforce the rigid body motion in each particle subdomain onto the overlapping fluid field. Zhang and co-workers \cite{ZHANG1,LIU1} proposed the immersed finite element method (IFEM) to use a flexible Lagrangian solid mesh that moves on top of a background Eulerian fluid mesh, which spans the entire computational domain. In IFEM the mesh generation is greatly simplified, and the continuity between the fluid and solid sub-domains is enforced by interpolating the velocities. Forces are distributed using the  reproducing kernel particle method (RKPM) delta function. In fact, the higher-ordered RKPM delta function enables non-uniform spatial meshes (on arbitrary geometries) for the fluid domain. Wang and Liu \citep{WANG20041305} proposed the extended immersed boundary method (EIBM), by including the submerged elastic solid occupying a finite volume in the flow domain. This was done by replacing the kinematic and dynamic matching of the fluid-solid interface and the effect of the immersed solid with nodal forces calculated in the context of finite element formulations. The equations in both the fluid and solid domains are approximated using the finite element method, while the continuity between the fluid and solid domains is enforced by interpolating the velocities and the distribution of forces delta function with the reproducing kernel particle method (RKPM). 

Finite cell method \cite{DUSTER1} is an example of well-established IB method. In the finite cell method, the computational domain and the embedded domain are discretized using a finite number of computational cells (usually having simple shapes like squares or cubes) that cover the entire domain. The unknown field variables are smoothly extended beyond the cells, cut by the original boundary. Special integration techniques are employed to treat cells cut by the boundary. The finite  cell  method (FCM) belongs  to  the  fictitious domain class \cite{GLOW2,RAMIERE1} of embedded  domain methods \cite{Nei1,Lohner1,Lohner2}. The main idea of the fictitious domain methods is that they extend the physical domain of interest beyond the complex boundaries into a larger embedding domain discretized by a simple structured grid. In the finite cell method, the computational domain and the embedded domain are discretized using a finite number of computational cells. Special integration techniques are employed to treat cells cut by the boundary. The fictitious domain concept has been used in numerous immersed  boundary type  methods,  which  have  become increasingly popular over the last decade for the simulation of  large-scale  computational fluid mechanics  problems.  Some examples  include  the extended finite element method \cite{Suk1,Has1,LEGRAIN1}, isogeometric   immersed   boundary   methods \cite{CASQUERO,CASQUERO2,XU2,KAMENSKY1,KAMENSKY2},  or  embedded  domain  approaches  focusing on Lagrange  multipliers \cite{BURMAN1,Gers1}, Nitsche type methods \cite{Baiges1,Hautefeuille1} and  discontinuous  Galerkin  methods \cite{Bastian1,RANGARAJAN1}.

In the present study, we develop and use the voxelized finite element (FE) boundary condition-enforced immersed boundary (VBCE-IB) method, which is an extension of the BCE-IB method introduced in \citep{BOURANTAS2021105162}, to numerically solve blood flow in arteries (internal flow). The flow domain is discretized using voxels/cubes that envelope the immersed object and it is no longer a box domain. The main advantage of the VBCE-IB method over the BCE-IB method is that a large percentage of the Eulerian grid nodes are located inside the immersed boundary
Additionally, the VBCE-IB method accurately satisfies both the governing equations and boundary conditions using velocity and pressure correction procedures. The velocity correction applies implicitly such that the velocity on the immersed boundary (Lagrangian points) interpolated from the corrected velocity values computed on the mesh nodes (Eulerian nodes) accurately satisfies the prescribed velocity boundary conditions. The incremental pressure correction scheme (IPCS) is a modified version of the projection method, which provides improved accuracy at little extra computational cost. We implement our method using FEniCS \citep{unknown_2019_fenics} (an open-source software package that solves partial differential equations) to numerically solve the incompressible Navier–Stokes equations. FEniCS encompasses software to specify variational forms of PDEs in the high-level Unified Form Language (UFL) and compiles these descriptions into numerical routines. This software has been shown to achieve high levels of computational performance, even in problems with billions of degrees of freedom \citep{RICHARDSON201932}. The proposed IB scheme applies to both external and internal flows, but in the present study we are particularly interested in internal flow problems. Our numerical scheme increases the computational efficiency of the IB method for internal flow cases by reducing the number of unused elements through a sophisticated and efficient mesh refinement inside the immersed boundary. Reducing the computational cost is of paramount importance for IB methods applied to internal flows because not all of the advantageous features of IB methods that have made them popular for external flow simulations apply to internal flow simulations. 

The remainder of the paper is organized as follows. In Section 2, we present the numerical formulation of the proposed immersed boundary (IB) method. In Section 3, we solve a benchmark problem (Poiseuille flow) to verify the accuracy of the proposed IB method. Section 4 presents numerical examples that demonstrate the accuracy and applicability of the proposed IB method. Section 5 contains conclusions.

\section{Numerical method}
We consider the incompressible viscous flow in a spatial domain $\Omega$, which contains an immersed boundary in the form of a closed surface $\partial S$. The immersed boundary is modeled as localized body forces acting on the surrounding fluid. The incompressible N-S equations—in their primitive variables (velocity $\bm{u}$ and pressure $p$) formulation—accounting for the immersed boundary are written as:

\begin{equation}
    \label{eq:momentum}
        \frac{\partial\bm{u}}{\partial t}
        + \left( \bm{u} \cdot \bm{\nabla}\right) \bm{u}  = 
    - \bm{\nabla} p 
    + \bm{\nabla} \cdot 2 \nu_{f} \bm{\varepsilon}(\bm{u}) 
    + \bm{f},
\end{equation}
\begin{equation}
    \label{eq:incompressibility-constraint}
    \bm{\nabla} \cdot \bm{u} = 0,
\end{equation}
subject to the velocity boundary condition on \(\partial S\)
\begin{equation}
    \label{eq:noslip_bc}
    \bm{u}(\bm{X}(s,t),t) = \bm{U}_{B},
\end{equation}
with \(\bm{U}_{B}\) being the prescribed velocity of the immersed boundary, and
\(v_{f}\) the kinematic viscosity of the fluid. The term
\(\bm{\varepsilon}(\bm{u})\) is the strain-rate tensor
defined as:
\begin{equation}
    \bm{\varepsilon}(\bm{u}) = \frac{1}{2}\left(
        \bm{\nabla}\bm{u} + \left( \bm{\nabla}\bm{u} \right)^\mathrm{T}
    \right).
\end{equation}
The forcing term \(\bm{f}\) is added to the right-hand side (RHS) of
the momentum Eq.\eqref{eq:momentum} to account for the presence of the
immersed boundary.
The forcing term
\(\bm{f}\left( \bm{x},t \right)\) is the local body force
density at the fluid nodes (Eulerian nodes), and it is obtained from
the surface force density \(\bm{F}\left( s,t \right)\)
(\(s\) is a parametrization of the immersed boundary surface) at the
immersed boundary (Lagrangian points---we use traditional terminology used in \citep{wu_shu_2009_implicit}). The forcing term is
expressed as:
\begin{equation}
    \bm{f}(\bm{x},t) = 
    \int_{\partial S} \bm{F}(s,t) \, \delta\left( \bm{x}-\bm{X}(s,t) \right)\mathrm{d}s,
\end{equation}
where \(\bm{x}\) and \(\bm{X}\left( s,t \right)\)
denote the Eulerian nodes and the Lagrangian point coordinates that discretize the
fluid domain and represent the immersed boundary, respectively. Eulerian nodes and
Lagrangian points interact through the Dirac delta function
\(\delta\left( \bm{x} - \bm{X}\left( s,t \right) \right)\). Therefore,
the velocity at the immersed boundary points is calculated from the velocity at the Eulerian nodes using the Dirac delta function
\begin{equation}
    \bm{U}(\bm{X}(s,t)) =
    \int_{\Omega} \bm{u}(\bm{x}) \, \delta\left( \bm{x}-\bm{X}(s,t) \right) \mathrm{d}V.
\end{equation}

\subsection{Incremental Pressure Correction Scheme (IPCS)}
\label{sec:ipcs}

The Incremental Pressure Correction Scheme (IPCS) \citep{goda_1979_multistep} is an operator splitting method that decouples the pressure and velocity fields of the N-S equations. IPCS is a modified version of the fractional
step method proposed by
\citet{chorin_1968_numerical}
and
\citet{temam_1969_approximation}. It improves the accuracy of the original scheme with little extra cost.
In our description, we consider no external forces for the N-S equations,
and we explain in detail how the IB forces were incorporated. The spatial discretization of velocity and pressure was performed using linear Lagrange $(P_{1}/P_{1})$ elements \cite{KHAN, BATHE1}. Since we used a splitting approach (IPCS method), the Ladyzhenskaya-Babuska-Brezzi (LBB) condition is inherently satisfied, and thus stabilization of the spatial discretization was not required \cite{BATHE1}. 

The IPCS scheme involves three steps. In the first step, we compute a tentative velocity \(\bm{u}^{*}\) by advancing the linear momentum Eq.\eqref{eq:momentum} in time using the backward Euler difference scheme, \(\dot{\bm{u}} = \frac{\bm{u}^{n+1} - \bm{u}^{n}}{\Delta t}\). We linearize the nonlinear terms using a semi-implicit method, such that the advection term \(\bm{u} \cdot \bm{\bm{\nabla}}\bm{u}\) becomes
\(\bm{u}^{n} \cdot \bm{\bm{\nabla}}\bm{u}^{n+1}\). Using this
semi-implicit approach for linearization, the Courant-Friedrichs-Lewy (CFL) condition, which limits the time step according to the velocity and spatial discretization to ensure stability of the solution, becomes less restrictive.
Additionally, the kinematic viscosity is written as
\(v_{f} = v_{f}\left( \bm{u}^{n} \right)\), which simplifies the N-S
equations that are now written as a linearized set of equations known
as the Oseen equations:
\begin{equation}
    \label{eq:oseen}
    \bm{u}^{n+1}
    + \Delta t\bm{u}^{n} \cdot \bm{\nabla}\bm{u}^{n+1}
    - \Delta t\bm{\nabla} \cdot 2v_{f}\bm{\varepsilon}(\bm{u}^{n+1})
    + \Delta t\bm{\nabla} p^{n+1} 
    = \bm{u}^{n},
\end{equation}
\begin{equation}
    \bm{\nabla} \cdot \bm{u}^{n+1} = 0.
\end{equation}

In Eq.\eqref{eq:oseen}, \(p^{n+1}\) is still unknown.
Therefore, we will compute the
tentative velocity \(\bm{u}^{*}\)
(\(\bm{u}^{*} \approx \bm{u}^{n+1}\)), by replacing in Eq.\eqref{eq:oseen}
\(p^{n+1}\) with the known value \(p^{n}\). 
This makes Eq.\eqref{eq:oseen} easier to solve for the tentative velocity
\(\bm{u}^{*}\):

\begin{equation}
    \label{eq:tentative-velocity}
    \bm{u}^{*} 
    + \Delta t\,\bm{u}^{n} \cdot \bm{\nabla}\bm{u}^{n+1}
    - \Delta t\,\bm{\nabla} \cdot 2v_{f}\bm{\varepsilon}\left( \bm{u}^{*} \right)
    + \Delta t\,\bm{\nabla} p^{n} 
    = \bm{u}^{n}.
\end{equation}
Equation \eqref{eq:tentative-velocity} is numerically solved to compute the tentative velocity
\(\bm{u}^{*}\) without using the incompressibility constraint
Eq.\eqref{eq:incompressibility-constraint}.

In the second step, we use the tentative velocity
\(\bm{u}^{*}\) to compute the updated velocity \(\bm{u}^{n+1}\),
which is divergence-free (and should fulfill the incompressibility
constraint). We define a function for the velocity correction as
\(\bm{u}^{c} = \bm{u}^{n+1} - \bm{u}^{*}\), and after some
algebra (subtracting Eq.\eqref{eq:tentative-velocity} from Eq.\eqref{eq:oseen}) 
and using that
\(\bm{\nabla} \cdot \bm{u}^{n+1} = 0 \Rightarrow \ \bm{\nabla} \cdot \bm{u}^{c} = - \bm{\nabla} \cdot \bm{u}^{*}\)
(incompressibility constraint) we obtain a new set of flow equations for
\(\bm{u}^{c}\):
\begin{equation}
    \label{eq:uc1}
    \bm{u}^c 
    + \Delta t\,\bm{u}^{n} \cdot \bm{\nabla}\bm{u}^c
    - \Delta t\,\bm{\nabla} \cdot 2v_{f}\bm{\varepsilon}\left( \bm{u}^c \right)
    + \Delta t\,\bm{\nabla}\Phi^{n+1}
    = 0,
\end{equation}
\begin{equation}
    \label{eq:uc2}
    \bm{\nabla} \cdot \bm{u}^{c} = - \bm{\nabla} \cdot \bm{u}^{*},
\end{equation}
where \(\Phi^{n+1} = p^{n+1} - p^{n}\). 
Equation \eqref{eq:uc1} can be further simplified to
\(\bm{u}^{c}+\Delta\text{t\ }\bm{\bm{\nabla}}\Phi^{n+1}=\bm{0}\)
and the system of Eqs.(10)-(11) is reduced to an elliptic type (Poisson)
partial differential equation for the pressure difference \(\Phi^{n}\)
\begin{equation}
    \nabla^2 \Phi^{n+1} = \frac{\bm{\nabla} \cdot \bm{u}^*}{\Delta t},
\end{equation}
with boundary conditions being those applied to the original flow
problem.

In the third and final step, we compute the updated velocity
\(\bm{u}^{n+1}\) and pressure \(p^{n+1}\) through
\(p^{n+1} = \Phi^{n+1} + p^{n}\) and
\(\bm{u}^{n+1} = \bm{u}^{*} + \Delta t\bm{\nabla}\Phi^{n+1}\),
respectively. 

The algorithmic procedure for our implementation of the IPCS for solving N-S equations is summarized below:

\begin{enumerate}
\item
  Calculate the tentative velocity \(\bm{u}^{*}\) by solving Eq.\eqref{eq:tentative-velocity}
\item
  Solve the Poisson equation
  \(\bm{}\nabla^2 \Phi^{n+1}\bm{=}\frac{\nabla \cdot \bm{u}^{*}}{\Delta t}\)
  (Eq.(12)) to obtain $\Phi^{n+1}$
\item
  Calculate the corrected velocity
  \(\bm{u}^{n+1} = \bm{u}^{*} + \Delta t\bm{\nabla}\Phi^{n+1}\)
  and pressure \(p^{n+1} = \Phi^{n+1} + p^{n}.\)
\end{enumerate}

\subsection{Boundary Condition Enforced Immersed Boundary (BCE-IB) method}
\label{sec:bce-ibm}

In this section, we describe the boundary condition enforced immersed
boundary (BCE-IB) method for the numerical solution of the incompressible 3D Navier--Stokes equations. The immersed boundary method has two steps:

\begin{itemize}
\item
  \emph{Predictor step}: where we numerically solve the N-S equations to
  compute the predicted velocity field
  \(\tilde{\bm{u}}\left( \bm{x} \right)\) by disregarding
  the body force terms in Eq.(1) (\(\bm{f} = \bm{0}\))
\end{itemize}

\begin{equation}
   \frac{\tilde{\bm{u}}-\bm{u}^{n}}{\Delta t}
   + \left( \bm{u}^n \cdot \bm{\nabla}\bm{u}^n \right) 
   = + \Delta t\bm{\nabla} p^{n} 
   - \Delta t\bm{\nabla} \cdot 2v_{f}\bm{\varepsilon}(\bm{u}^{n})
\end{equation}

\begin{itemize}
\item
  \emph{Corrector step}: where we account for the effect of body forces
  and update the predicted velocity field \(\tilde{\bm{u}}\) to
  the updated one \(\bm{u}^{n + 1}\)\textbf{,} which satisfies the
  boundary condition
  \(\bm{u}\left( \bm{X}\left( s,t \right),t \right) = \bm{U}_{B}\)
\end{itemize}

\begin{equation}
     \frac{\bm{u}^{n + 1} - \tilde{\bm{u}}}{\Delta t} 
    = \bm{f}^{n + 1}.
\end{equation}

In the predictor step, Eq.(13) is used to compute the predicted velocity
field \(\tilde{\bm{u}}\) under the incompressibility constraint
(mass conservation), which couples the velocity and pressure. The
predictor step accounts for the first of the three steps
in the IPCS method. In this step, we compute the predicted velocity
\(\tilde{\bm{u}}\), which in this case is the tentative velocity
\(\bm{u}^{*}\) (Step 1 in the IPCS algorithmic procedure). The
corrector step involves the evaluation of the unknown body force
\(\bm{f}^{n + 1}\) and the update of the predicted velocity
\(\tilde{\bm{u}}\) to the physical one
\(\bm{u}^{n + 1}\) (Steps 2 and 3 in the IPCS algorithmic
procedure).

Accurate and efficient computation of the unknown body force \(\bm{f}^{n + 1}\) is an important
advantage of the Boundary Condition Enforced Immersed Boundary (BCE-IB) method. In this method,
the body force \(\bm{f}^{n + 1}\) is equivalent to a velocity
correction, which applies implicitly such that the velocity
\(\bm{U}\left(\bm{X}\left( s,t \right),t \right)\) on the
Lagrangian points, interpolated from the physical velocity
\(\bm{u}\left( \bm{x},t \right)\) computed on the Eulerian nodes, equals
to the prescribed boundary velocity \(\bm{U}_{B}\) (Eq.~\eqref{eq:noslip_bc}). The
body force term \(\bm{f}^{n + 1}\) is computed using the equation
\begin{equation}
    \bm{U}_B^{n+1}\left( \bm{X}^{n+1} \right) =
    \int_{\Omega} \left( \tilde{\bm{u}} + \Delta t\frac{\bm{f}^{n + 1}}{\rho} \right)
    \delta\left( \bm{x} - \bm{X}^{n+1} \right) \mathrm{d}V,
\end{equation}
derived by substituting Eq.(14) to Eq.(6). The force density
\(\bm{f}^{n + 1}\left( \bm{x},t \right)\), is evaluated on the
Eulerian nodes, and it is computed by distributing (`spreading') the
boundary force \(\bm{F}\left( \bm{X}\left( s,t \right),t \right)\)
on the Lagrangian points through the Dirac delta function
\(\delta\left( \bm{x - X}\left( s,t \right) \right)\) (see
Eq.(5)). Eq.(15) is written as
\begin{equation}
    \bm{U}_B^{n+1} \left( \bm{X}^{n+1} \right) = 
    \int \left( \tilde{\bm{u}} + \Delta t\frac{\int {\bm{F}^{n+1} 
    \left( \bm{X}^{n + 1} \right)
    \delta\left( \bm{x} - \bm{X}^{n+1} \right)\mathrm{d}s}}{\rho} \right)
    \delta\left( \bm{x} - \bm{X}^{n+1} \right)\mathrm{d}V,
\end{equation}
and \(\bm{U}_{B}^{n + 1}\) is no longer correlated with
\(\bm{f}^{n + 1}\) but instead with \(\bm{F}^{n + 1}\).

To compute the boundary force \(\bm{F}^{n + 1}\), we rewrite Eq.(16)
in a discrete form that results in an algebraic system of equations.
We represent the immersed boundary by a set of Lagrangian
points
\(\bm{X}_{i} = \left( X_{i},Y_{i},Z_{i} \right),\ i = 1,2,\ldots,M\)
and the flow domain by the Eulerian nodes
\(\bm{x}_{j} = \left( x_{j},y_{j},z_{j} \right),\ j = 1,2,\ldots,N\).
In our analysis,
the Eulerian mesh in the vicinity of the immersed object
has a uniform Cartesian grid structure
with grid spacing \(h\).
Furthermore, the Dirac function
\(\delta\left( \bm{x} - \bm{X}\left( s,t \right) \right)\) is
approximated by a continuous kernel distribution
\(D\left( \bm{x}_{i} - \bm{X}^{j} \right)\) given as
\begin{equation}
    D_{ij} = D\left( \bm{x}_i - \bm{X}^j \right) =
    \left( \frac{1}{h}\delta\left( \frac{x_i - X^j}{h} \right) \right)
    \left( \frac{1}{h}\delta\left( \frac{y_i - Y^j}{h} \right) \right)
    \left( \frac{1}{h}\delta\left( \frac{z_i - Z^j}{h} \right) \right),
\end{equation}
with the kernel \(\delta\left( r \right)\) proposed by
\citet{lai_peskin_2000_immersed}
\begin{equation}
    \delta(r) = 
        \begin{cases}
        \begin{alignedat}{2}
            \frac{1}{8}\left( 3 - 2|r| + \sqrt{1 + 4 |r| - 4r^2} \right),   &\quad&     |r| & \leq 1 \\
            \frac{1}{8}\left( 5 - 2 |r| - \sqrt{-7 + 12|r| - 4r^2} \right), &\quad& 1 < |r| & \leq 2 \\
            0,                                                              &\quad&     |r| & > 2
        \end{alignedat}
        \end{cases}
\end{equation}

Using Eq.(5) for the force term,
\begin{equation}
    \bm{f}^{n+1}(\bm{x}_j) =
    \sum_{i=1}^M {\bm{F}^{n+1}\left( \bm{X}_i^{n+1} \right)
    D_h^{ij}\Delta S_i},
\end{equation}
Eq.(16) is written as
\begin{equation}
   \bm{U}_{B}^{n+1}(\bm{X}_i^{n+1}) 
   = \sum_{j=1}^N \tilde{\bm{u}}(\bm{x}_j) D_h^{ij} h^3 
   + \sum_{j=1}^N \sum_{k=1}^M \frac{\bm{F}^{n+1}(\bm{X}_i^{n+1}) \Delta t}{\rho}
   D_h^{kj} D_h^{ij} h^3,
\end{equation}
where \(\Delta S_{i}\) is the area of the \emph{i\textsuperscript{th}}
surface element (segment). Eq.(20) forms a well-defined system of
equations for the variables
\(\bm{F}_i^{n + 1}\bm{\ }\left( i = 1,2,\ldots,M \right)\).
Eq.(20) can be written in matrix notation as
\begin{equation}
    \bm{A}_{\bm{F}}\bm{F} = \bm{B}_{\bm{F}},
\end{equation}
with \(\bm{A}_{\bm{F}},\bm{F}\) and
\(\bm{B}_{\bm{F}}\) defined as
\begin{equation}
    \bm{A}_{\bm{F}} =
    \frac{\Delta t}{\rho}h^3
    \begin{bmatrix}
        D_{11} \Delta S_1 & D_{12} \Delta S_1 & \dots  & D_{1N} \Delta S_1 \\
        D_{21} \Delta S_2 & D_{22} \Delta S_2 & \dots  & D_{2N} \Delta S_2 \\
        \vdots            & \vdots            & \ddots & \vdots            \\
        D_{M1} \Delta S_M & D_{M2} \Delta S_M & \dots  & D_{MN} \Delta S_M
    \end{bmatrix}
    \begin{bmatrix}
        D_{11} & D_{12} & \dots  & D_{1M} \\
        D_{21} & D_{22} & \dots  & D_{2M} \\
        \vdots & \vdots & \ddots & \vdots \\
        D_{N1} & D_{N2} & \dots  & D_{NM}
    \end{bmatrix},
\end{equation}
\begin{equation}
    \bm{B}_{\bm{F}} =
    \bm{U} - h^3 \bm{D}^\mathrm{T} \tilde{\bm{u}} =
    \begin{bmatrix}
        \bm{U}_1 \\
        \bm{U}_2 \\
        \vdots       \\
        \bm{U}_M 
    \end{bmatrix}
    - h^3
    \begin{bmatrix}
        D_{11} & D_{21} & \dots  & D_{N1} \\
        D_{12} & D_{22} & \dots  & D_{N2} \\
        \vdots & \vdots & \ddots & \vdots \\
        D_{1M} & D_{2M} & \dots  & D_{NM}
    \end{bmatrix}
    \begin{bmatrix}
        \tilde{\bm{u}}_1 \\
        \tilde{\bm{u}}_2 \\
        \vdots               \\
        \tilde{\bm{u}}_N
    \end{bmatrix},
\end{equation}

\begin{equation}
    \bm{F} =
    \begin{bmatrix}
        \bm{F}_1 \\
        \bm{F}_2 \\
        \vdots       \\
        \bm{F}_M 
    \end{bmatrix},
\end{equation}
where \(\bm{U}_i\ (i\bm{=}1,2,\ldots,M\bm{)}\),
\(\bm{F}_i\bm{\ (}i\bm{=}1,2,\ldots,M\bm{)}\)
and
\({\tilde{\bm{u}}}_j\bm{\ }\left( j = 1,2,\ldots,M \right)\)
are the abbreviations for
\(\bm{U}_{B}^{n + 1}\left( \bm{X}_{\bm{i}}^{n + 1} \right)\),
\(\bm{F}^{n + 1}\left( \bm{X}_{\bm{i}}^{n + 1} \right)\) and
\(\tilde{\bm{u}}\left( \bm{x}_{j} \right)\), respectively.
By solving the system of equations (Eq.(22)) using a direct solver, we obtain the unknown boundary force
\(\bm{F}_{\bm{i}}^{n + 1}\bm{\ }\left( i = 1,2,\ldots,M \right)\)
at all Lagrangian points. Boundary forces
\(\bm{F}_{\bm{i}}^{n + 1}\) are then substituted into Eq.(19)
and Eq.(16) to calculate the body force \(\bm{f}^{n + 1}\) on the Eulerian nodes, and the
corrected physical velocity \(\bm{u}^{n + 1}\)\emph{. }

\subsection{Algorithmic procedure}

Our approach combines the IPCS (section~\ref{sec:ipcs})
with the BCE-IB method (section~\ref{sec:bce-ibm}).
The velocity field predicted by IPCS 
is corrected to account for the force contribution
from the immersed boundary.
The proposed solution procedure can be summarized as follows;
to update the solution from time instance \(n\) to \(n+1\):


\begin{enumerate}

\item
  Calculate the tentative velocity \(\tilde{\bm{u}}\) using Eq.(13)
\item
  Compute the matrix \(\bm{A}_{\bm{F}}\) using Eq.(22)
\item
  Solve the system (Eq.(21)) to compute the boundary force
  \(\bm{F}^{n + 1}\left( \bm{X}_i^{n + 1} \right)\bm{,\ (}i\bm{=}1,2,\ldots,M\bm{)}\)
  on the Lagrangian points. Substitute the computed boundary force
  \(\bm{F}^{n + 1}\left( \bm{X}_i^{n + 1} \right)\)
  into Eq.(19) to obtain the body force
  \(\bm{f}^{n + 1}\left( \bm{x}_{j}^{n + 1} \right)\bm{,\ (}j\bm{=}1,2,\ldots,N\bm{)}\) on the Eulerian points.
\item
  Use \(\tilde{\bm{u}}\) to update the corrected velocity
  \(\bm{u}^{(n+1)}=\Delta t \bm{f}^{n+1} + \tilde{\bm{u}}\)
  using Eq.(14)
 \item
  Solve the Poisson equation
  \(\bm{}\nabla^2 \Phi^{n+1}\bm{=}\frac{\bm{\nabla} \cdot \bm{u}^{n+1}}{\Delta t}\) for $\Phi^{n}$
  (Eq.(12))
\item
  Recompute the updated velocity
  \(\bm{u}^{n+1} = \bm{u}^{n} + \Delta t\bm{\bm{\nabla}}\Phi^{n+1}\)
  and pressure \(p^{n+1} = \Phi^{n+1} + p^{n}.\)
\item
  Set physical/updated velocity \(\bm{u}^{\left( n + 1 \right)}\) as \(\bm{u}^{\left( n  \right)}\) and repeat steps (1) to (4) until the desired
  solution is achieved.
\end{enumerate}

Steps 1, 5, 6 apply to the IPCS method without the presence of the immersed boundary, while Steps 2, 3, 4 calculate the velocity field (on the Eulerian nodes) due to the presence of the immersed boundary.

\section{Algorithm verification}
In this section, we demonstrate the accuracy of the proposed method in internal flow cases. We consider two benchmark flow problems, Hagen-Poiseuille flow in a straight tube and flow in a U-bend geometry. The numerical results obtained were compared against the analytical solution for the Hagen-Poiseuille flow example and against experimental data in the U-bend geometry flow case. 

\subsection{Poiseuille flow in a cylindrical tube}

In the first benchmark flow example, we consider the Hagen-Poiseuille flow in a straight cylindrical tube. The driving force of the flow is the pressure difference applied to the inlet and outlet. In our simulations, we set the density of the fluid to \(\rho = 1,050~\mathrm{kg\,m^{-3}}\) and the dynamic viscosity to \(\mu = 0.00345~\mathrm{N\,s/m^{2}}\). The tube has length \(L = 10R\) and radius \(R = 0.005\) m, and is immersed in the voxelized flow domain, as shown in Fig.~1. We demonstrate the accuracy of the voxelized BCE-IB method by comparing our numerical findings with the analytical solution (seec\citep{BOURANTAS2021105162}). The immersed boundary (cylinder) is represented by a set of surface points (Lagrangian nodes). The voxelized flow domain is discretized first with hexahedral elements and second with tetrahedral elements (generated by splitting the hexahedral elements). At the inlet (red circle in Fig.~1b) and outlet (orange circle in Fig.~1b), we apply pressure boundary conditions, while at the remaining walls (surface yellow Fig.~1b), we apply no-slip velocity boundary conditions (\(\bm{u} = 0\)). 

\begin{figure}[ht]
    \centering
    \includegraphics[width=440pt,height=300pt]{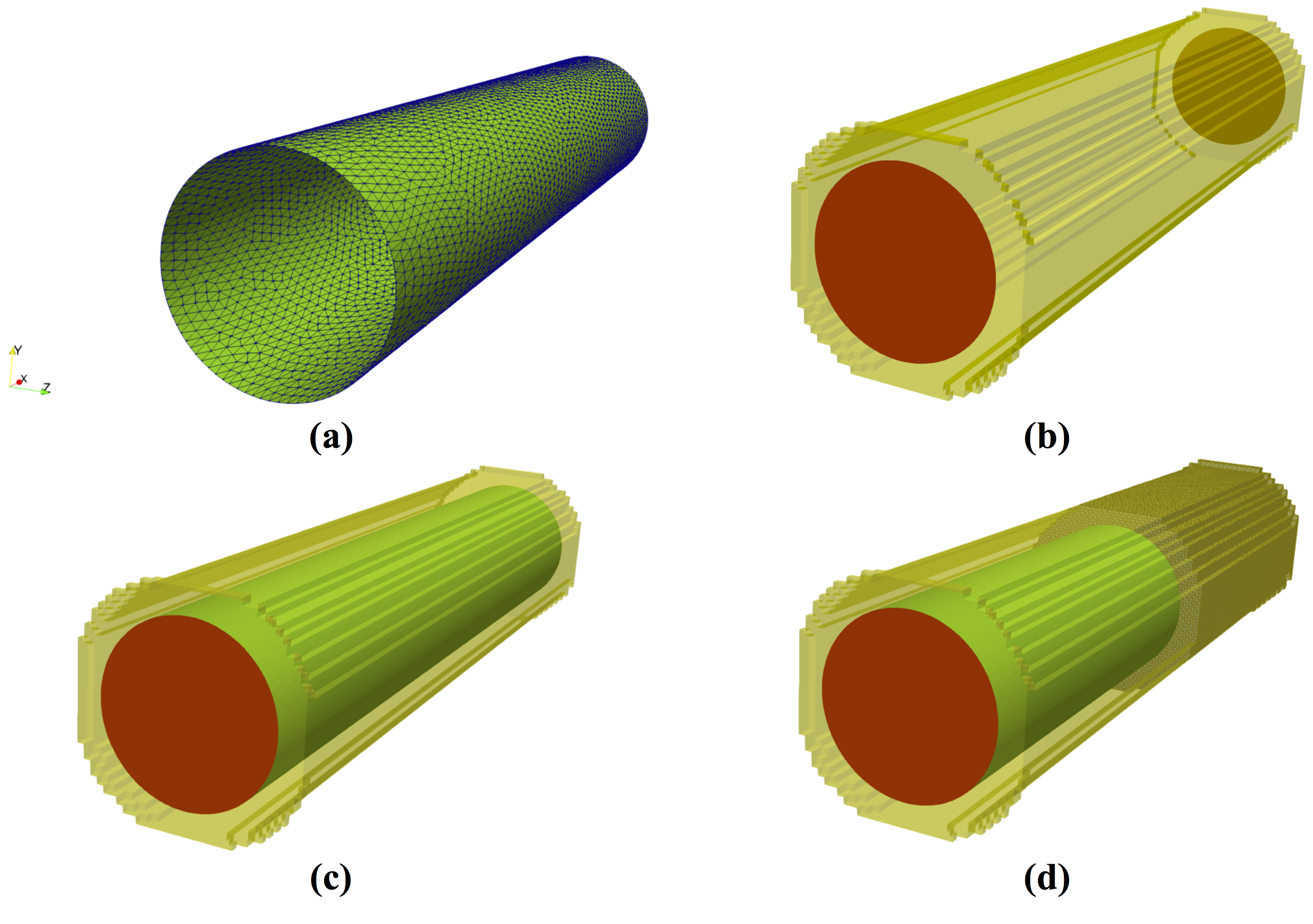}
    \caption{
        Poiseuille flow cylinder geometry
        \textbf{(a)}
        immersed boundary given as STL file with nodes and facets
        \textbf{(b)}
        flow domain for the Poiseuille flow example, with 
        inlet (circle in red color), 
        outlet (circle in orange color) and
        walls (yellow color) boundaries; 
        \textbf{(c)}
        immersed boundary into the voxelized flow domain, and
        \textbf{(d)}
        voxelized flow domain along with the cross-section of the mesh.
    }
    \label{fig:my_label}
\end{figure}

We do not conduct a mesh-independence analysis in order to ensure a mesh-independent numerical solution. Instead, we use a dense mesh for the voxelized flow domain. As previously demonstrated (see \citep{BOURANTAS2021105162}), we set the voxelized Eulerian mesh resolution based on the distribution of the Lagrangian nodes. We set the Eulerian mesh resolution $h$ using the mean value $\Delta S_{mean}$ of the surface area $\Delta S$ assigned to each Lagrangian point, such that $h = \sqrt[]{\Delta S_{mean}}$. Furthermore, we examine how the accuracy of the numerical results is affected (boundary effects) by the dimensions (radius) of the voxelized flow domain (the number of hexahedral elements (cubes) located outside the immersed boundary (see Fig. 2)). In our simulations we use five, six and seven rows of elements outside the immersed boundary (see Table 1). The time step was set to \(dt = 2.5 \times 10^{- 3}\) (this time step ensures the accuracy and stability of the numerical solution). The simulation ends when the normalized root mean square difference
\begin{equation}
    \text{NRMSE} = \frac{\sqrt{\frac{\sum_{i = 1}^{N}\left( u_{i}^{t + dt} - u_{i}^{t} \right)^{2}}{N^2}}}{\left( \max\left( u_{i}^{t} \right) - \min\left( u_{i}^{t} \right) \right)},
\end{equation}
between two successive time instances for all three velocity components is less than \(10^{-6}\) (practically, the flow reaches a steady state). The exact solution for the velocity field components is \(u_{y} = u_{z} = 0\) and \(u_{x}\left( r \right) = U_{\max}\left( 1 - \left( \frac{r}{R} \right)^{2} \right)\),
with \(U_{\max} = - \frac{R^{2}}{4\mu}\frac{\text{dp}}{\text{dz}}\)
being the maximum velocity. By setting the pressure difference to
\(\frac{\text{dp}}{\text{dz}} = \frac{1}{L}\) (applying
\(p_{\text{inlet}} = 1\) and \(p_{\text{outlet}} = 0\)) the Reynolds
number becomes \(\mathrm{Re} = \frac{\rho U_{\max}L}{\mu} \approx 551\).

We use an immersed boundary consisting of 4,869 nodes and 9,570 facets. The mean area of the facets is $\Delta S_{mean} = 2.5 \times 10^{-7}$ $m^{2}$, and the voxelized Eulerian mesh (flow domain) resolution is $h = \sqrt[]{\Delta S_{mean}} = 0.0005$ m. We discretize the voxelized flow domain using hexahedral elements. We use five, six and seven rows of elements outside the immersed boundary (Fig.2), which leads to 57,626 elements (63,761 nodes), 62,853 elements (69,264 nodes), and 67,570 elements (74,245 nodes), respectively. The numerical results computed using the different voxelized flow domains were interpolated on a body-fitted mesh (38,756 nodes and 213,913 elements) with tetrahedral elements (used only for visualization). 

Table 1 lists the NRMSE for the different voxelized flow domains (with hexahedral elements) compared with the analytical solution computed on the body-fitted tetrahedral mesh. The results show that the dimensions of the voxelized domain do not affect the accuracy of the numerical solution. To demonstrate the applicability of the proposed scheme, we used tetrahedral elements to discretize the flow domain. We use the same geometrical specifics used in the tetrahedral meshes above, and the hexahedral elements are now split into tetrahedral elements, generating tetrahedral meshes with 345,756 elements (63,761 nodes), 377,118 elements (69,264 nodes) and 405,420 elements (74,245 nodes), respectively. Table 2 lists the NRMSE for the different voxelized flow domains (with tetrahedral elements) compared with the analytical solution computed on the body-fitted tetrahedral mesh. The results show that the dimensions of the voxelized domain do not affect the accuracy of the numerical solution.

\begin{table}
    \centering
    \caption{Normalized root mean square error (NRMSE) of the voxelized immersed boundary method (with hexahedral elements) numerical results against the analytical solution omputed on the body-fitted tetrahedral mesh for the Poiseuille flow example. The three hexahedral meshes correspond to five, six and seven rows of elements outside the immersed boundary.
    }
    \begin{tabular*}{\textwidth}
    {@{\extracolsep{\fill}} ccccccc }
        \toprule
        mesh & $h$ & number of nodes & number of elements & $u_{x}$ - NRMSE   & $u_{y}$ - NRMSE  & $u_{z}$ - NRMSE \\
        \midrule
        $ 1$ & $5 \times 10^{-4}$ & 63,761 & 57,626 & $1.33 \times 10^{-2}$  & $5.52 \times 10^{-4}$ & $3.77 \times 10^{-4}$\\
        $2$ & $5 \times 10^{-4}$ & 69,264 & 62,853 & $2.42 \times 10^{-2}$ & $4.63 \times 10^{-4}$ & $3.56 \times 10^{-4}$ \\
        $3$ & $5 \times 10^{-4}$ & 74,245 & 67,570 & $1.96 \times 10^{-2}$  & $3.23 \times 10^{-4}$ & $2.18 \times 10^{-4}$ \\
        \bottomrule
    \end{tabular*}
    \label{tab:coronary-norms1}
\end{table}

\begin{table}
    \centering
    \caption{Normalized root mean square error (NRMSE) of the immersed boundary method (with tetrahedral elements) numerical results against the analytical solution omputed on the body-fitted tetrahedral mesh for the Poiseuille flow examplefor the Poiseuille flow example. The three tetrahedral meshes correspond to five, six and seven rows of elements outside the immersed boundary.
    }
    \begin{tabular*}{\textwidth}
    {@{\extracolsep{\fill}} ccccccc }
        \toprule
        mesh & $h$ & number of nodes & number of elements & $u_{x}$ - NRMSE  & $u_{y}$ - NRMSE & $u_{z}$ - NRMSE\\
        \midrule
        $ 1$ & $5 \times 10^{-4}$ & 63,761 & 345,756 & $1.29 \times 10^{-2}$  & $4.02 \times 10^{-4}$ & $3.77 \times 10^{-4}$\\
        $2$ & $5 \times 10^{-4}$ & 69,264 & 377,118 & $1.31 \times 10^{-2}$ & $3.12 \times 10^{-4}$ & $3.56 \times 10^{-4}$ \\
        $3$ & $5 \times 10^{-4}$ & 74,245 & 405,420 & $1.15 \times 10^{-2}$  & $2.96 \times 10^{-4}$ & $2.46 \times 10^{-4}$ \\
        \bottomrule
    \end{tabular*}
    \label{tab:coronary-norms2}
\end{table}

\begin{figure}[ht]
    \centering
    \includegraphics[width=470pt,height=120pt]{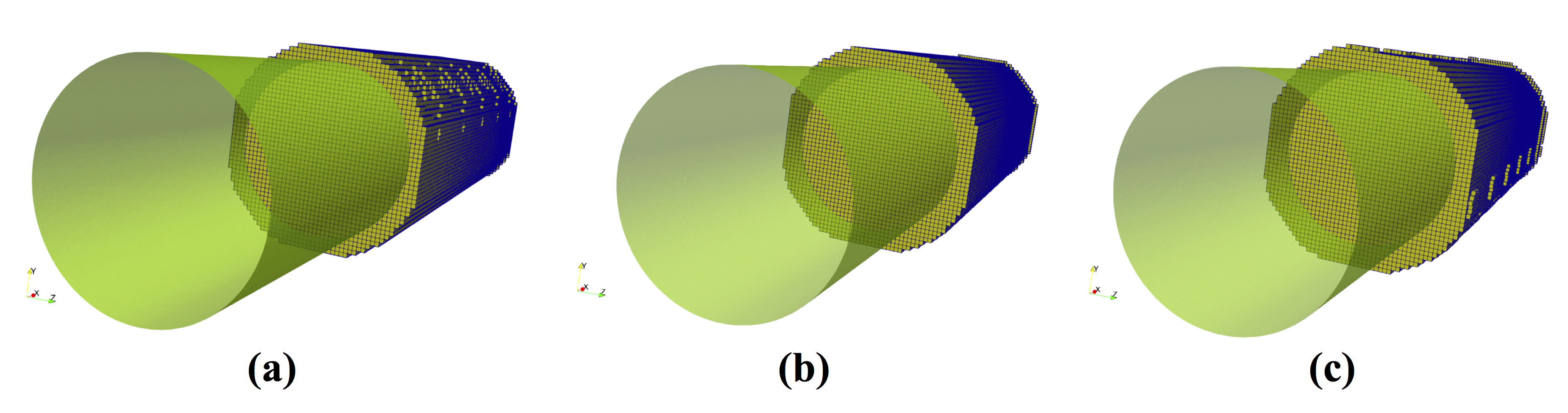}
    \caption{
        Poiseuille flow example; immersed boundary (green cylinder) and a cross-section of the voxelized mesh (hexahedral elements) with
        \textbf{(a)}  five
        \textbf{(b)} six, and
        \textbf{(c)} seven elements 
        outside the immersed boundary.
    }
    \label{fig:my_label}
\end{figure}

Figure 3 shows the velocity streamlines for \(\mathrm{Re} = 551\) produced on the tetrahedral body-fitted mesh. The velocity field computed with the IB method on the Eulerian nodes has been projected on the visualization mesh for visualization purposes (the visualization mesh is not used in the computation of the velocity and pressure field). 

\begin{figure}[ht]
    \centering
    \includegraphics[width=280pt,height=130pt]{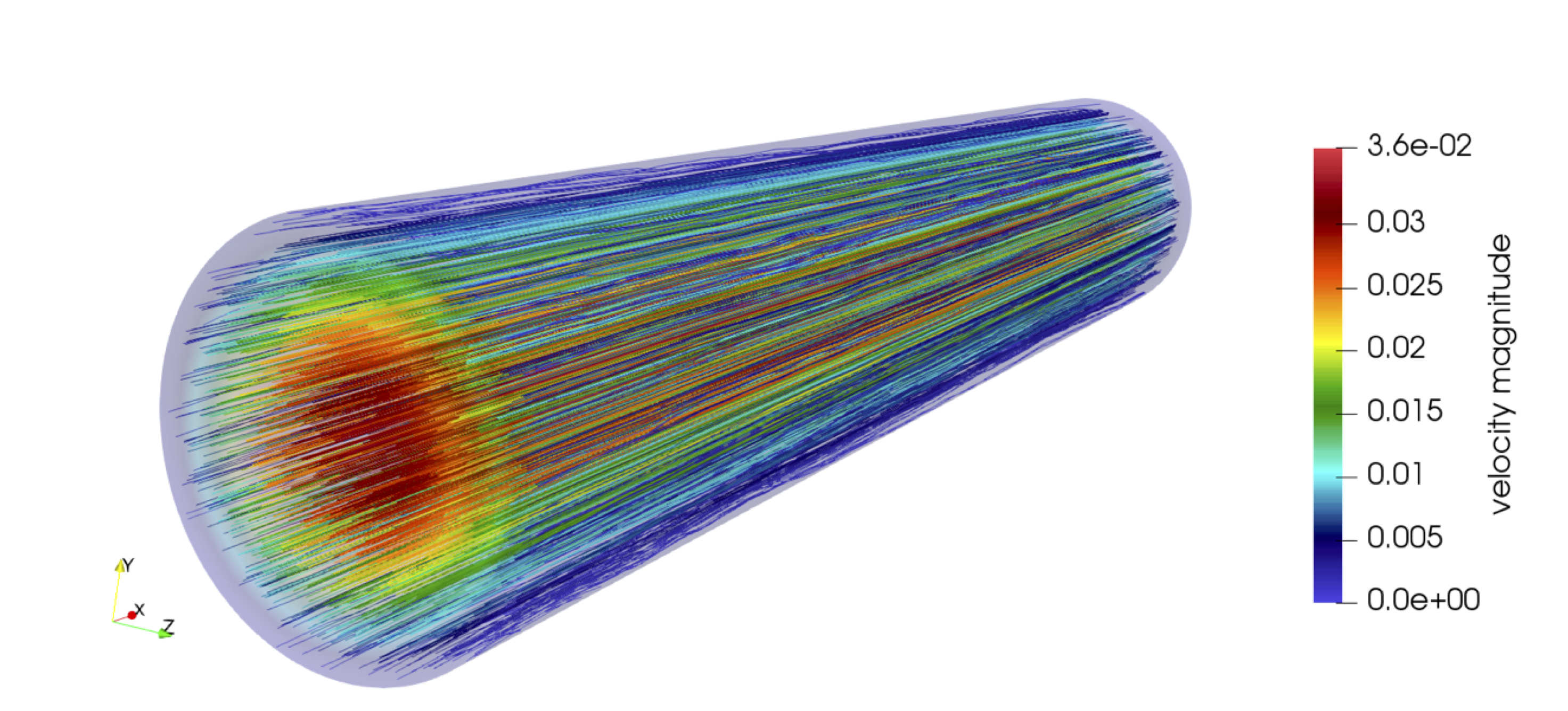}
    \caption{
        Velocity streamlines for \(\mathrm{Re} = 551\) using a visualization mesh of  213,913 linear ($P_{1}/P_{1}$) elements for the Poiseuille flow example. The maximum velocity computed for \(\mathrm{Re} = 551\) was 0.0361 (the theoretical value is 0.0362).
    }
    \label{fig:my_label}
\end{figure}

\subsection{3D angle tube bend}

In the second benchmark problem, we demonstrate the accuracy of the proposed method considering the flow in a three-dimensional 90° angle tube bend, as shown in Fig.4. With this flow example we demonstrate that the proposed numerical IB method can accurately simulate flows in tortuous tubes where secondary flows can occur. 

For the U-bend flow problem, experimental flow data are available. \citet{vandevosse_etal_1989_finite} performed laser Doppler velocimetry experiments to obtain the center-plane axial velocities at a set of different angles around the tube bend. To reduce flow disturbances due to inflow boundary conditions, we extended the inlet and outlet by $R_{D} = 4$ mm (the length of the inner diameter \(D_{i}\)). In our numerical simulations, we set the Reynolds number --- computed based on the tube inner diameter \(D_{i}\) and the mean inlet velocity \(\overline{U}\) --- to \(\mathrm{Re} = 300\), to match the experimental conditions. The inner diameter \(D_{i}\) and the curvature radius \(R\) of the tube were set to 4 and
24 mm, respectively.

\begin{figure}[ht]
    \centering
    \includegraphics[width=480pt,height=210pt]{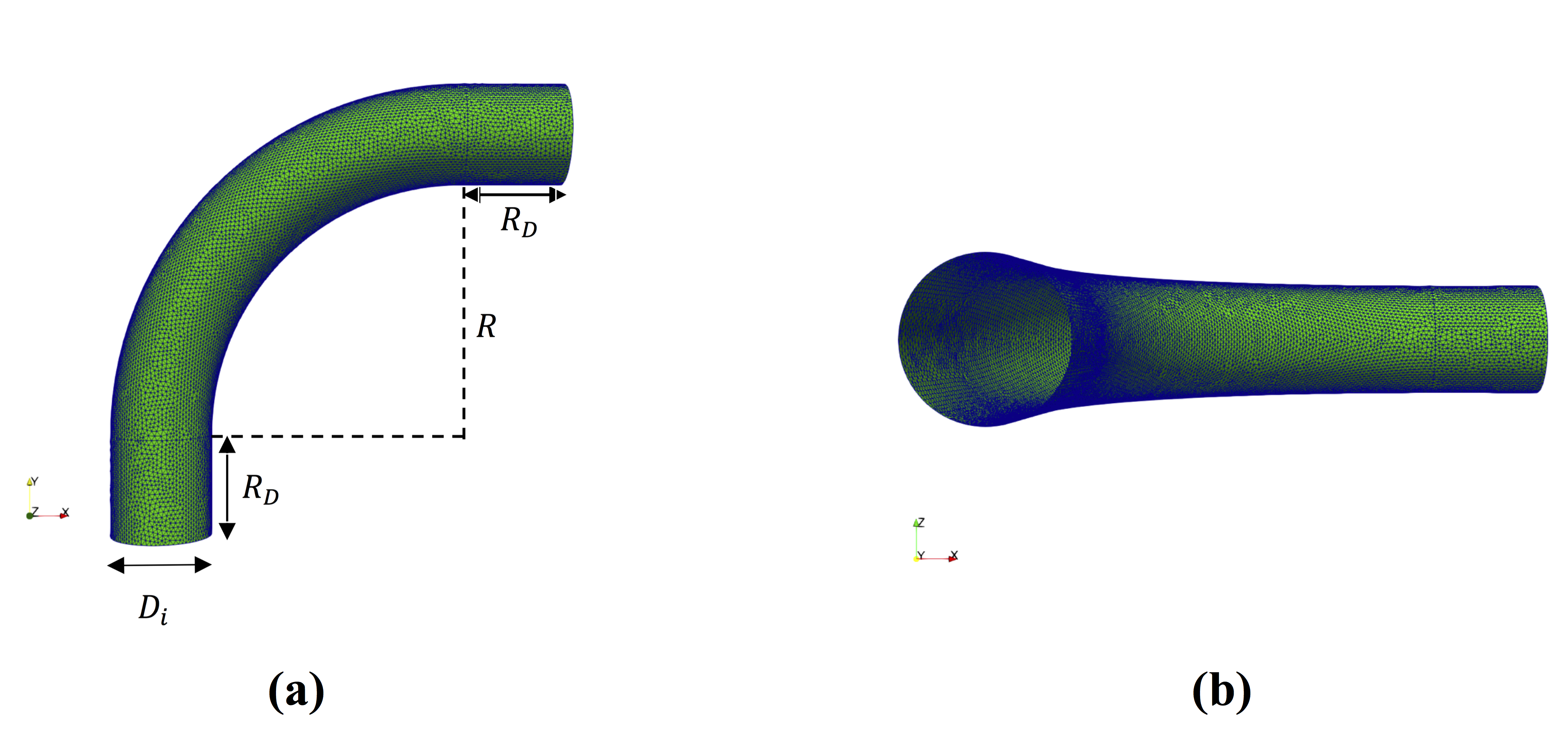}
    \caption{
        U-bend geometry
        \textbf{(a)} top and
        \textbf{(b)} side view.
        The points on the surface define the Lagrangian points of the immersed boundary $R_{D}=1 \ mm$, $R=24 \ mm$, $D_{i}=4 \ mm$.
    }
    \label{fig:my_label}
\end{figure}

\begin{figure}
    \centering
    \includegraphics[width=430pt,height=180pt]{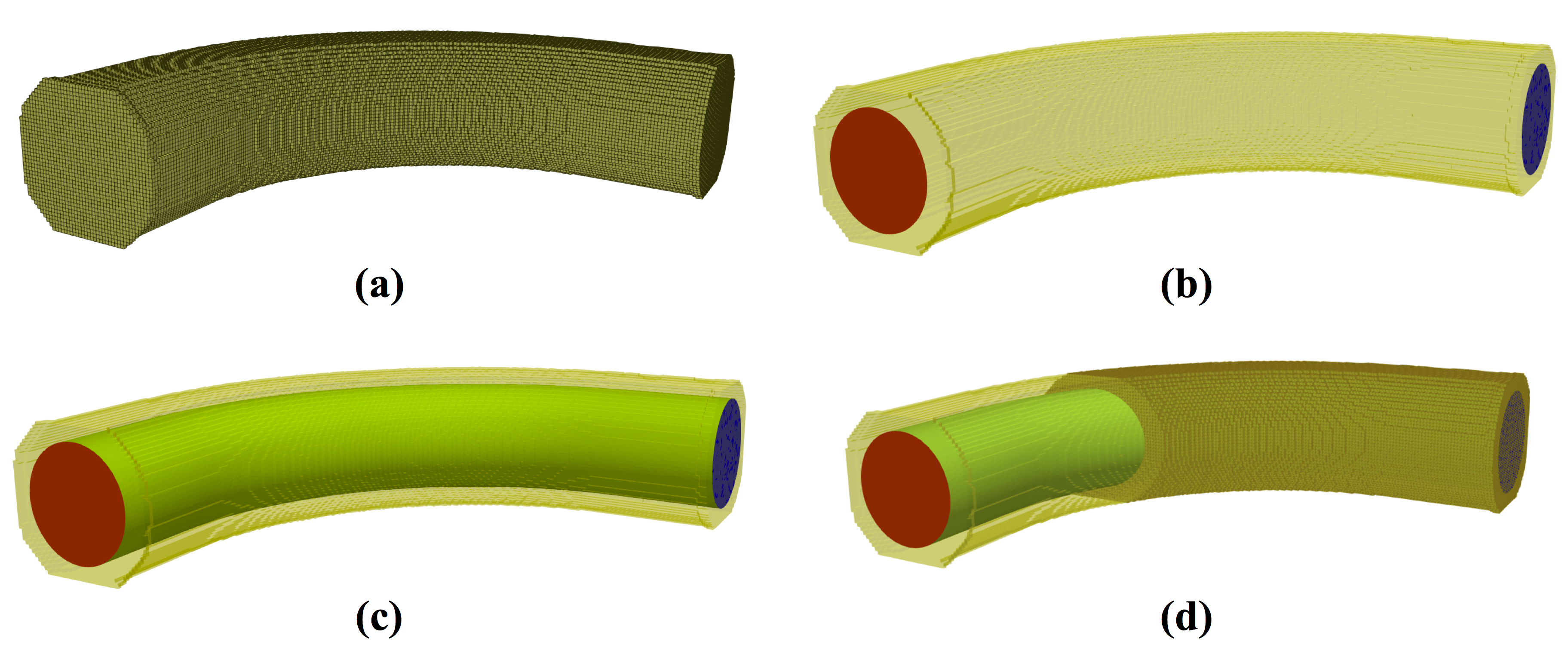}
    \caption{
        U-bend flow example
        \textbf{(a)}
        Immersed boundary given as STL file with nodes and facets
        \textbf{(b)}
        Flow domain for the Poiseuille flow example, with 
        inlet (circle in red color), 
        outlet (circle in blue color) and
        walls (yellow color) boundaries; 
        \textbf{(c)}
        Immersed boundary into the voxelized flow domain, and
        \textbf{(d)}
        voxelized flow domain along with the cross-section of the tetrahedral mesh.
    }
    \label{fig:my_label}
\end{figure}

The U-bend geometry (immersed object) is embedded within the voxelized flow domain (Fig.5a). We use a parabolic velocity profile (\(u_y(r) = U_{\max}\left(1 - \left( \frac{r}{R} \right)^{2} \right)\),  \(U_{\max} = 0.122625~\mathrm{m/s}\)) at the inlet (red circle in Fig.5b), while at the outlet (blue circle in Fig.5b) we apply zero pressure \(\left( p = 0 \right)\). At the remaining boundaries (shown in yellow in Fig.5b) of the flow domain, we set no-slip boundary conditions. 

We do not apply a mesh-independence analysis study  to ensure a mesh-independent numerical solution. Instead, we use a dense mesh to discretize the voxelized flow domain. The resolution $h$ of the uniform voxelized mesh (Eulerian mesh) is based on the average area $\Delta S$ (and consequently on the average length) of the surface elements (facets) of the immersed boundary. We use tetrahedral elements to discretize the flow domain. We use an immersed boundary consisting of 15,810 nodes and 31,462 facets. The mean area of the facets is $\Delta S_{mean} = 2.5 \times 10^{-7}$ m$^{2}$ and therefore the voxelized Eulerian mesh (flow domain) resolution is $h = \sqrt[]{\Delta S_{mean}} = 0.0005$ m. We use six rows of elements outside the immersed boundary. The voxelized flow domain consists of 3,162,114 linear elements ($P_{1}/P_{1}$) and 558,401 nodes. The numerical results computed using the voxelized flow domain were interpolated on a body-fitted mesh (used only for visualization having 143,406 nodes and 808,984 tetrahedral elements). We compared the results obtained using body-fitted mesh and voxelized flow domain. The NRMSE is $ 2 \times 10^{-3}$ (maximum value for all three velocity components), highlighting that the proposed method with the voxelized flow domain is accurate, and decreases the computational cost for the IB method for internal flows.

Fig.6 shows a comparison of the axial velocity computed using the IB method, and the experimental data (red dots) at the tube outlet. The axial velocity is computed on the center plane at the start of the outlet extension (see Fig.4a), where excellent agreement between the velocity predicted using the proposed IB method and experimental results is observed. The numerical results obtained using the linear elements (\(P_{1}/P_{1}\))  are also in good agreement with the experimental data, highlighting the accuracy of the proposed method.

\begin{figure}[ht]
    \centering
    \includegraphics[width=280pt,height=210pt]{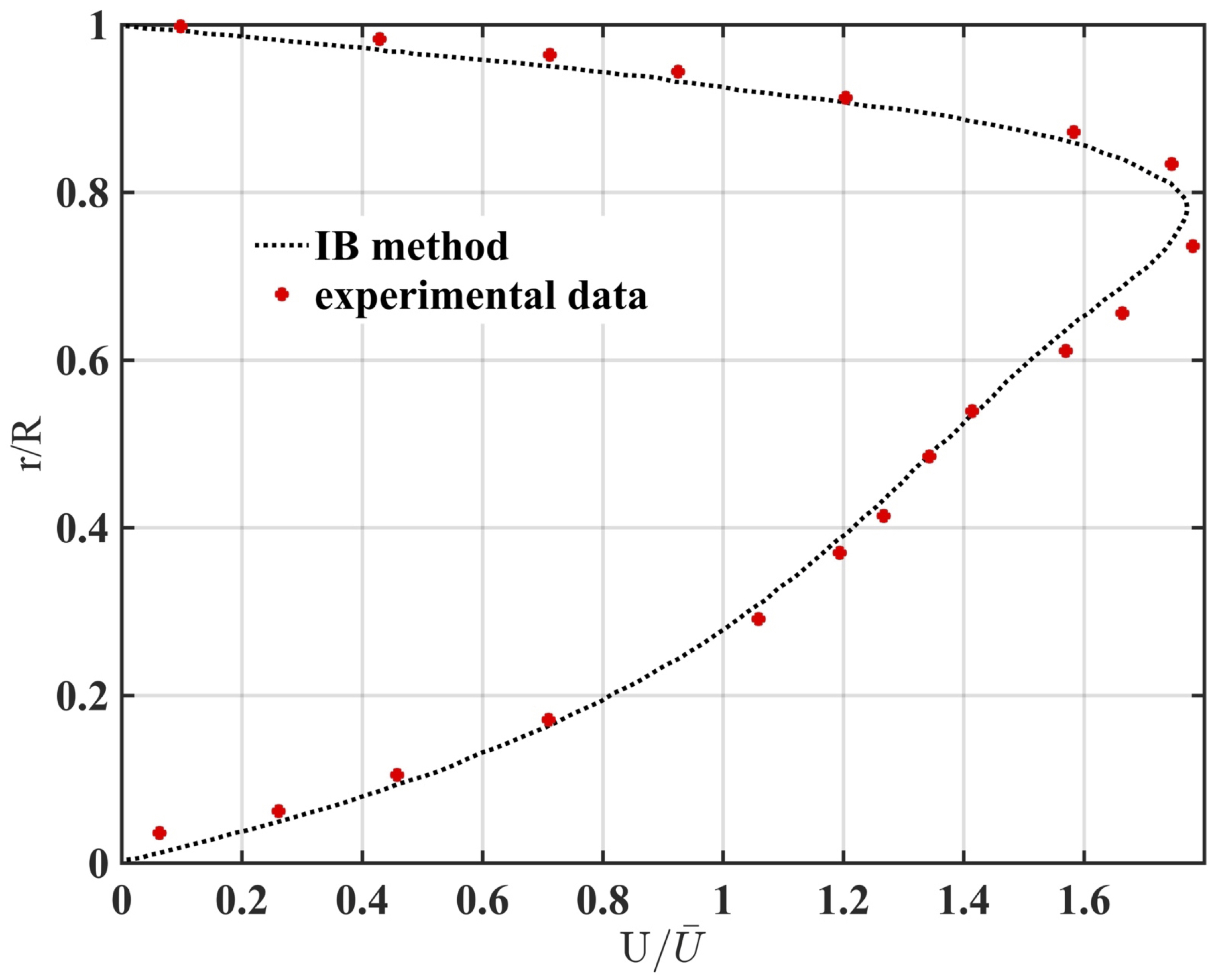}
    \caption{
        Axial velocity computed with the proposed IB method (dashed line) 
        compared with the experimental data (red dots) \citet{vandevosse_etal_1989_finite} at the tube outlet for the U-bend flow problem.
    }
    \label{fig:my_label}
\end{figure}

\section{Numerical results - Applications to complex flow domains}
In this section, we demonstrate the accuracy and applicability of the proposed method. We highlight the use of a voxelized geometry as the flow domain in terms of robustness, and we demonstrate the applicability of our method in internal flow problems with complex geometries. In the numerical examples considered in this section, we chose to use linear $(P_{1}/P_{1})$ tetrahedral elements (instead of hexahedral elements) to discretize the voxelized flow domain. Even if the number of tetrahedral elements is higher than the number of hexahedral elements (for the same voxelized flow domain), local refinement using tetrahedral elements is more straightforward (there are no hanging nodes) compared to hexahedral elements.

\subsection{Flow in a blood vessel bifurcation}
In this example, we simulate the blood flow in a  bifurcation (Fig.7a). The bifurcation geometry (immersed boundary) is represented by a triangle mesh and given as an STL file. The inlet and two outlets of the vessel geometry were extruded to ensure that the flow is fully developed. The vessel has length $L = 0.0296$ m, while the inlet has a mean radius of $R_{in}= 0.0034$ m and the two outlets of $R_{out} = 0.0023$ m. The voxelized flow domain (Fig.7b) envelopes the immersed boundary, as shown in Fig.7c. 

\begin{figure}[ht]
    \centering
    \includegraphics[width=440pt,height=390pt]{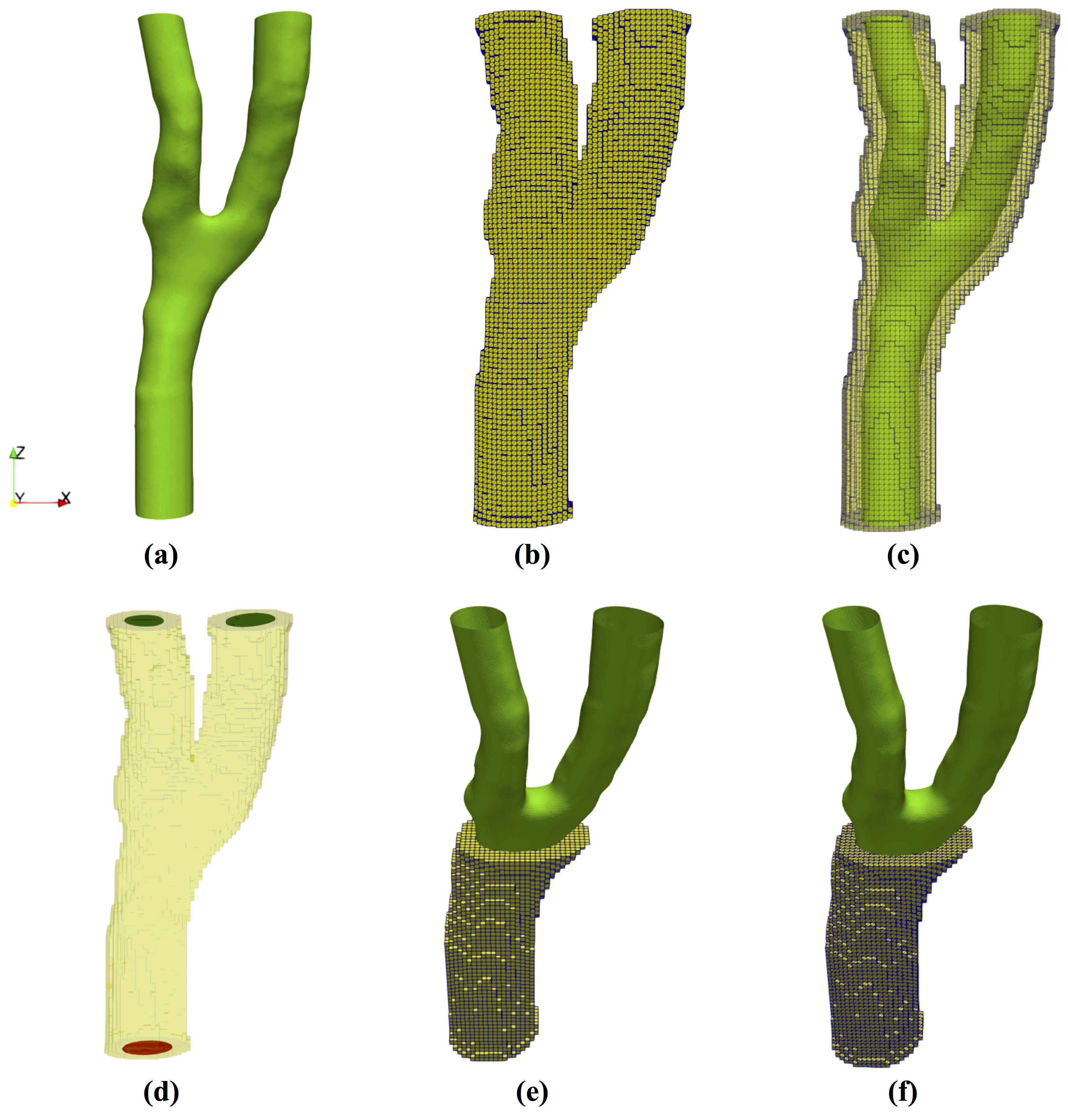}
    \caption{
        Flow in a vessel bifurcation example
        \textbf{(a)}
        immersed boundary given as STL file 
        \textbf{(b)}
        voxelized flow domain
        \textbf{(c)}
        immersed boundary enveloped into the voxelized flow domain
        \textbf{(d)}
        flow domain for the Poiseuille flow example, with 
        inlet (circle in red color), 
        outlet (circle in blue color) and
        walls (yellow color) boundaries
        \textbf{(e)}
        voxelized flow domain along with the cross-section of the hexahedral mesh
        \textbf{(f)}
        voxelized flow domain along with the cross-section of the tetrahedral mesh.
    }
    \label{fig:my_label}
\end{figure}

We do not apply a mesh-independence analysis to ensure a mesh-independent numerical solution. Instead, we use a dense mesh for the voxelized flow domain. We used 4,601,112 linear $(P_{1}/P_{1})$ tetrahedral elements (816,570 nodes) to discretize the voxelized flow domain (the voxelized mesh was generated using an in-house Python code). The dense mesh of 4,601,112 tetrahedral elements ensures a grid-independent numerical solution (in a previous study, a mesh of 627,040 tetrahedral elements ensured a grid-independent solution). The resolution $h$ of the uniform voxelized mesh (Eulerian mesh) is based on the average area $\Delta S_{mean}$ (and consequently on the average length) of the surface elements (facets) of the immersed boundary. We used a resolution of $h = 0.0005$ m, which corresponds to $\Delta S_{mean} = 2.5 \times 10^{-7}$ m$^{2}$, and seven elements outside the immersed boundary, to eliminate flow instabilities near the flow domain boundaries due to the presence of the immersed boundary. The mesh generation is a straightforward and efficient procedure since it takes less than 60 seconds to generate the voxelized mesh with 4,601,112 linear tetrahedral elements using a laptop computer with an i7 quad-core processor and 16 GB of internal memory.

We set the total time for the simulation to $T = 3.2$ sec (four cardiac cycles), and the time step to $dt = 5 \times 10^{-3}$ sec. At the inlet (red area located at the bottom surface in Fig.7d), we apply the velocity waveform shown in Fig.8 (we use constant velocity with components only in the $z-$ axis), and zero pressure boundary conditions at the two outlets (green area on the top surface in Fig.7d). At the remaining walls (yellow surface in Fig.7d), we apply no-slip boundary conditions. We use the Newtonian model for blood flow, with dynamic viscosity $\mu = 0.00345 Pa s$ and density of $\rho = 1,056 \frac{kg}{m^{3}}$. To demonstrate the accuracy of our IB method, we compare the results obtained using the current method with those for a body-fitted mesh with tetrahedral elements. The body-fitted mesh consists of 154,681 nodes and 860,818 linear tetrahedral elements \(\left( P_{1}/P_{1} \right)\). 

Table 3 lists the normalized root mean square error
\begin{equation}
\text{NRMSE} = \frac{\sqrt{\frac{\sum_{i = 1}^{N}\left( u_{i}^{IB} - u_{i}^{body fitted} \right)^{2}}{N^2}}}{\left( \max\left( u_{i}^{body fitted} \right) - \min\left( u_{i}^{body fitted} \right) \right)}
\end{equation}
at different time instances. The results obtained using our IB method and body-fitted mesh reported in Table 3 are, for practical purposes, indistinguishable as in the bioengineering applications differences of less than 5$\%$ would be overwhelmed by uncertainties in biomechanical properties. The decisive advantage of our method is the ease of generating patient-specific computational grids with practically no increase in computational cost. The only computational overhead of our IB method in comparison to the approach relying on a body-fitted mesh is due to the numerical solution of the linear system given by Eq.(21). For the computation of flow in the coronary artery bifurcation conducted here, this overhead was negligible (less than 1$\%$ of the total computation time). Fig.9 shows the streamlines in the coronary artery bifurcation at time \(t = 2.7\) sec and \(0.25\) sec (peak systole).

\begin{figure}[ht]
    \centering
    \includegraphics[width=440pt,height=170pt]{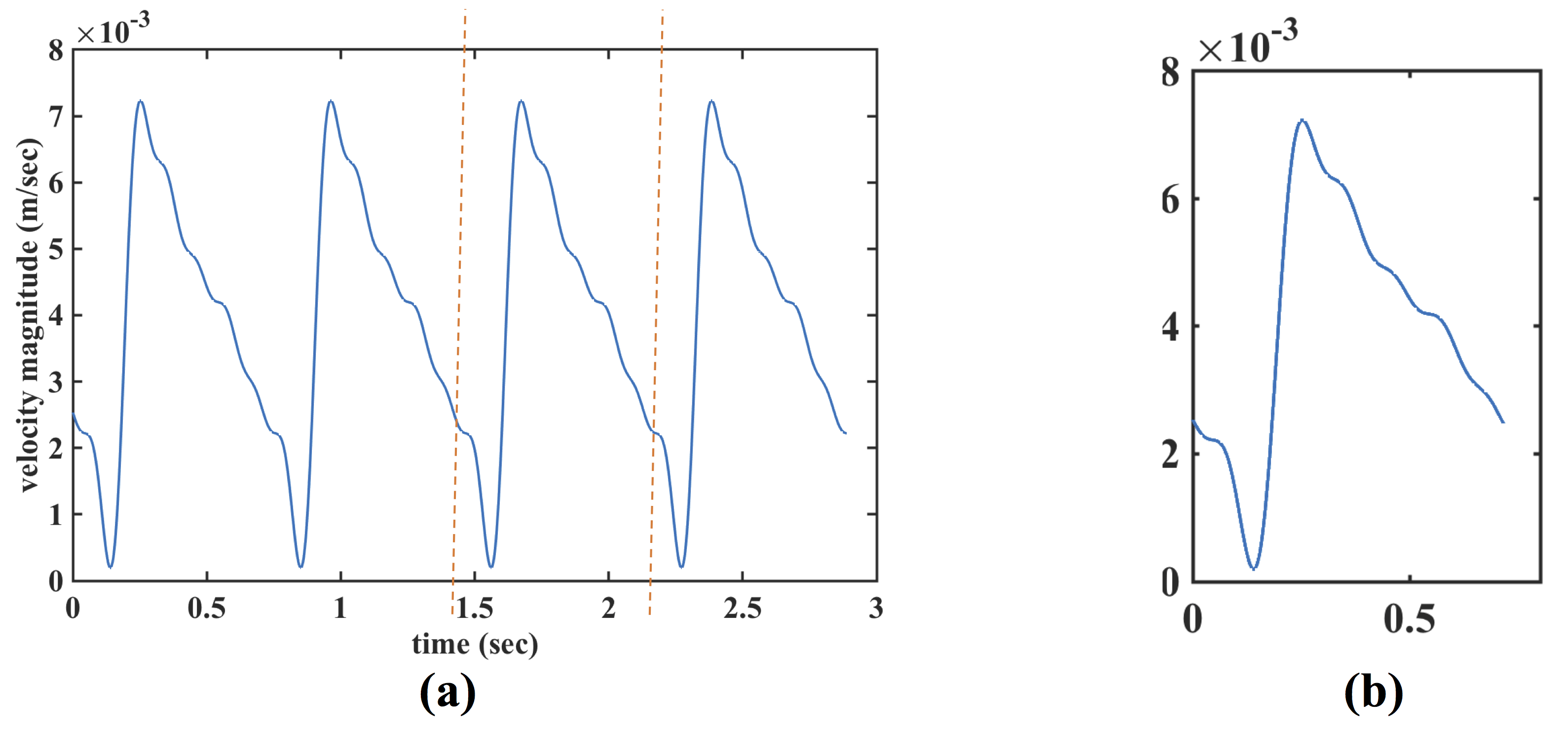}
    \caption{
    \textbf{(a)} Pulsatile velocity waveform (shown for three cardiac cycles) and \textbf{(b)} the pulsatile velocity waveform (one cardiac cycle) imposed at the inlet for the flow in a vessel bifurcation example.
    }
    \label{fig:my_label}
\end{figure}

\begin{table}
    \centering
    \caption{NRMSE of the voxelized flow domain using tetrahedral and hexahedral elements at different time instances for the flow in a vessel bifurcation example (see Fig.7 and Fig.9).
    }
    \begin{tabular*}{\textwidth}
    {@{\extracolsep{\fill}} ccccccccc }
        \toprule
        velocity component & $t=2.5 s$ & $t=2.6 s$ & $t=2.7 s$ & $t=2.8 s$ & $t=2.9 s$ & $t=3.0 s$ & $t=3.1 s$ & $t=3.2 s$\\
        \midrule
        $u_x$ & 2.29 & 3.78 & 1.78  & 3.18 & 4.27  & 2.51 & 2.25 & 2.22  \\
        $u_y$ & 3.05 & 4.04 & 1.93  & 3.71  & 3.97  & 3.16 & 3.64 & 3.88  \\
        $u_z$ & 4.53 & 4.18 & 3.60  & 2.82  & 3.78  & 4.27 & 3.33 & 2.82  \\
        \bottomrule
    \end{tabular*}
    \label{tab:coronary-norms3}
\end{table}

\begin{figure}[ht]
    \centering
    \includegraphics[width=370pt,height=480pt]{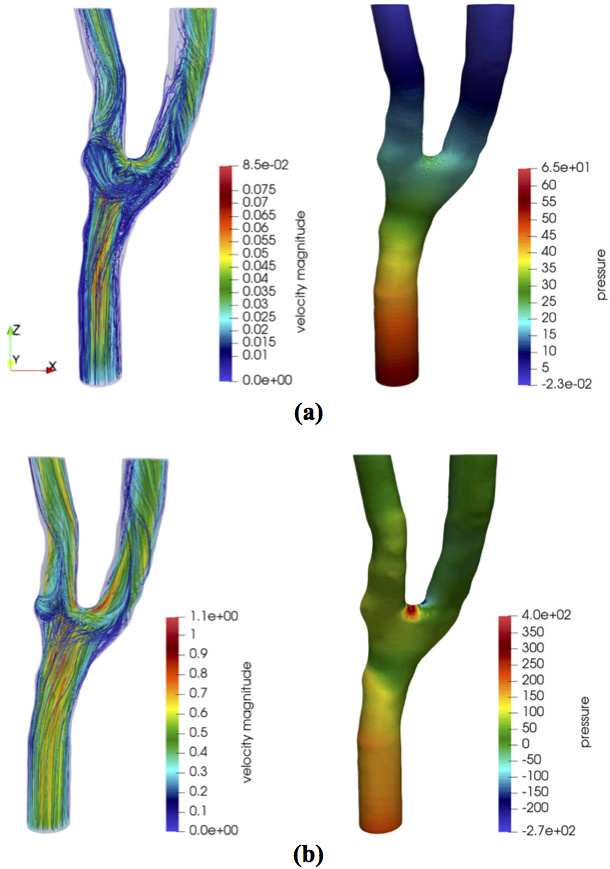}
    \caption{
        Streamlines (left) and 
        pressure (right) at
        \textbf{(a)}
        peak systole pressure and
        \textbf{(b)}
        diastolic minimum for the coronary artery bifurcation flow example.
    }
    \label{fig:my_label}
\end{figure}

\subsection{Flow in the descending aorta with aneurysm}
In this example, we simulate blood flow in the descending aorta with an aneurysm shown in Fig.10a. The immersed boundary (descending aorta) is 25 cm long in the $z$- axis and is embedded within a voxelized geometry that envelopes the immersed boundary (Fig.10c). 

\begin{figure}[ht]
    \centering
    \includegraphics[width=420pt,height=420pt]{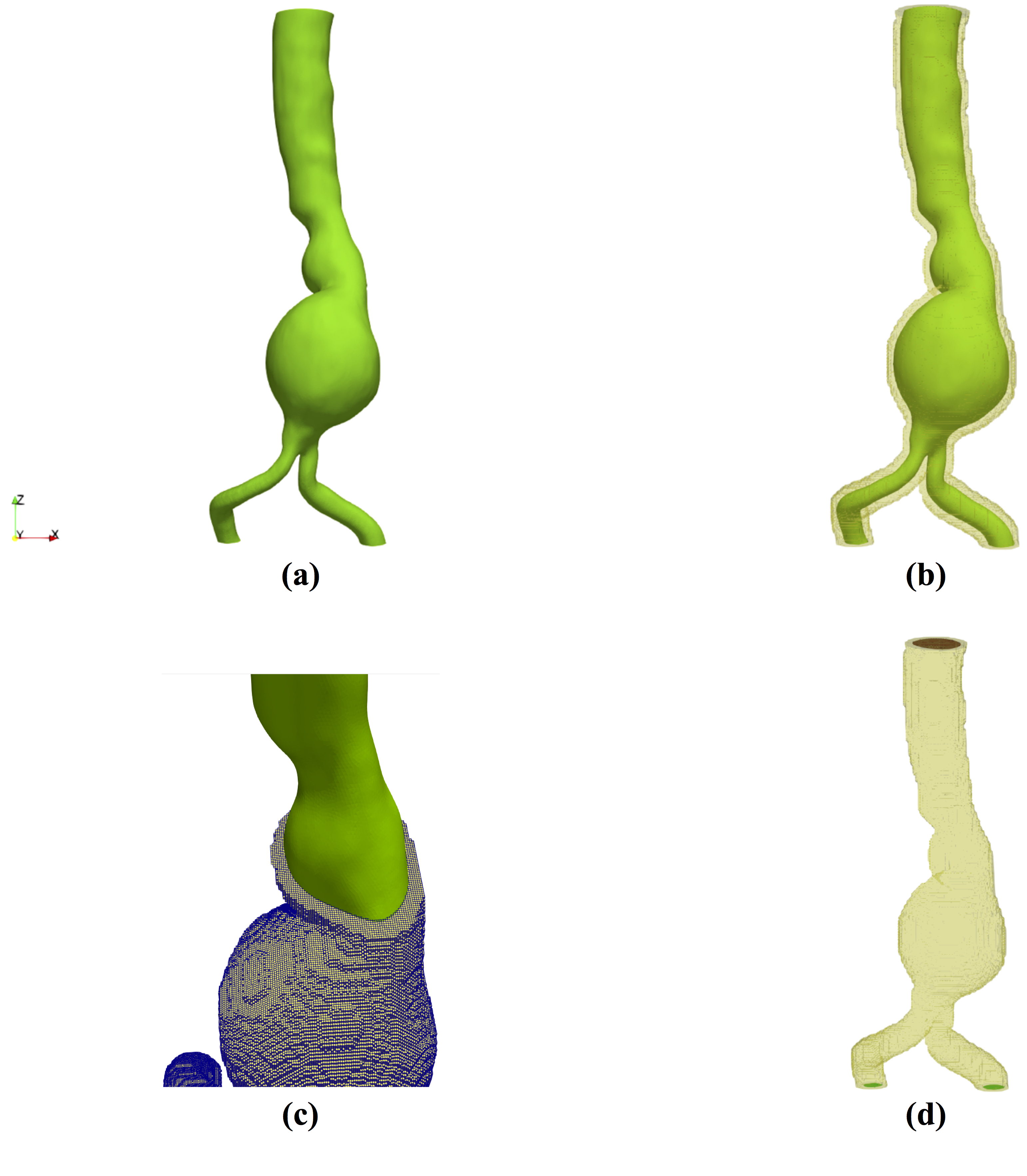}
    \caption{
        Flow in the descending aorta with aneurysm example
        \textbf{(a)}
        immersed boundary given as STL file 
        \textbf{(b)}
        voxelized flow domain (yellow color)
        \textbf{(c)}
        immersed boundary along with the cross-section of the tetrahedral mesh
        \textbf{(d)}
        flow domain for the descending aorta with aneurysm flow example, with 
        inlet (circle in red color), 
        outlet (circle in green color) and
        walls (yellow color) boundaries.
    }
    \label{fig:my_label}
\end{figure}

The voxelized flow domain is discretized using linear \(\left( P_{1}/P_{1} \right)\) tetrahedral elements, that envelopes the immersed boundary. The immersed boundary consists of 116,218 nodes and 232,127 triangular surface elements (facets). The mesh resolution $h$ of the voxelized flow domain is computed based on the average area $\Delta S_{mean}$ of the facets of the immersed boundary. The mean area of the facets is $\Delta S_{mean} = 1.06 \times 10^{-7}$ m$^{2}$ and therefore, the mesh resolution of the voxelized flow domain is $h = \sqrt[]{\Delta S_{mean}} = 0.00035$ m. We use six rows of elements outside the immersed boundary. We discretize the flow domain using 12,718,206 ($P_{1}/P_{1}$) elements and 3,558,401 nodes. This mesh, which ensures a grid-independent numerical solution, was generated in less than $30$ sec.

\begin{figure}[ht]
    \centering
    \includegraphics[width=420pt,height=180pt]{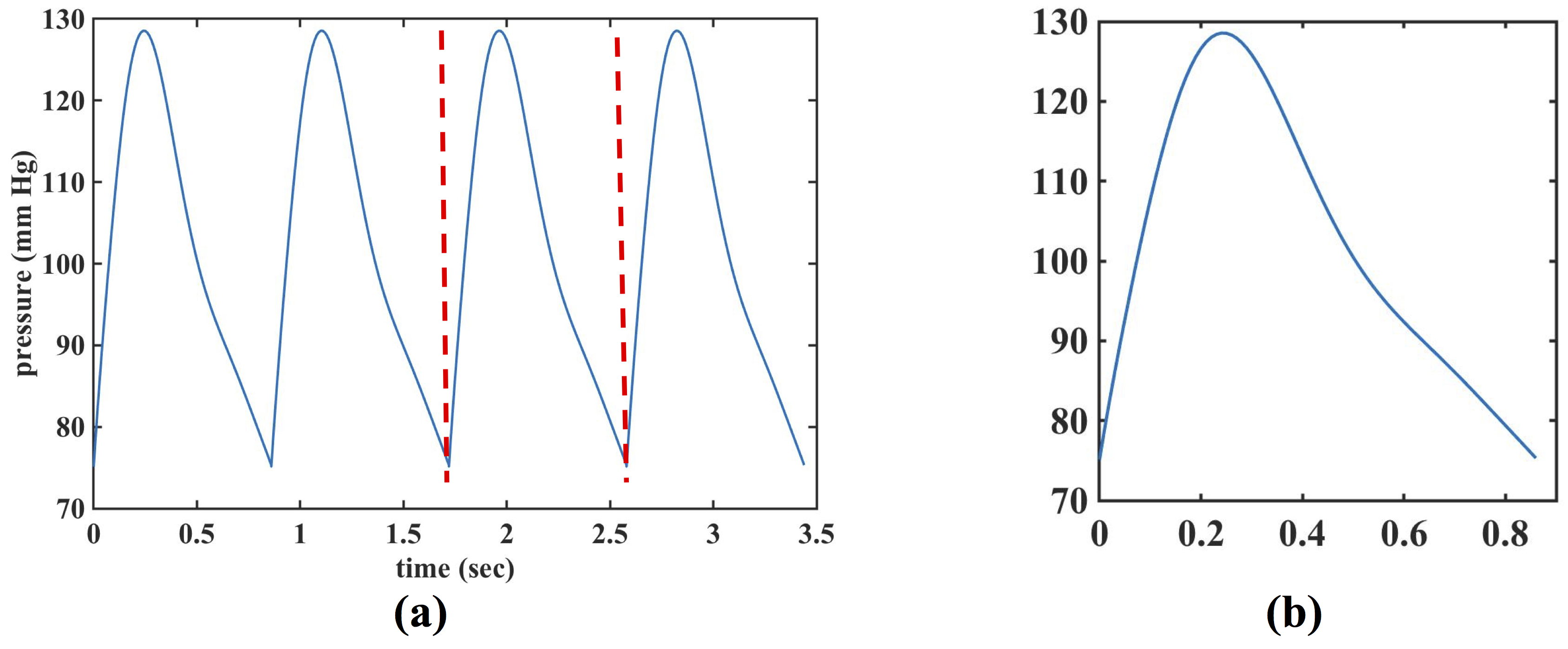}
    \caption{
        \textbf{(a)} Pulsatile pressure waveform (shown for four cardiac cycles) and \textbf{(b)} the pulsatile pressure waveform (one cardiac cycle) imposed at the inlet for the flow in the descending aorta with aneurysm example.
    }
    \label{fig:my_label}
\end{figure}

We simulate four cardiac cycles, and we set the total time of the simulation to $T = 3.44$ sec and the time step to $dt = 1 \times 10^{-4}$ sec. We apply the time-dependent pressure profile shown in Fig.11 at the inlet (red area located at the top surface in Fig. 10d) and zero pressure boundary conditions at the two outlets (green areas on the top surface in Fig.10d). At the remaining walls (yellow surface in Fig.10d), we apply no-slip boundary conditions. We use the Newtonian model for blood flow, with dynamic viscosity $\mu = 0.00345 Pa s$ and density of $\rho = 1,056 \frac{kg}{m^{3}}$. We visualize the results for the voxelized immersed boundary method simulation by projecting the numerical solution computed on the voxelized mesh to a body-fitted mesh with tetrahedral elements (this mesh is only for visualization and is not used in the simulation). We use the visualization body-fitted tetrahedral mesh (in total 4,018,371 elements) to numerically solve the flow equations using the same boundary conditions, and we compare the numerical solution (velocity field components) of the voxelized immersed boundary solver projected on the body-fitted mesh tetrahedral mesh nodes with the numerical solution obtained using the body-fitted mesh tetrahedral mesh at different time instances in the cardiac cycle. Table 4 lists the normalized root mean square error (NRMSE) between the two solutions (IB projected and body-fitted) on the body-fitted tetrahedral mesh vertices at different time instances. 
The results obtained using our IB method and body-fitted mesh reported in Table 4 are, for practical purposes, indistinguishable as in the bioengineering applications; differences of less than 5$\%$ would be overwhelmed by uncertainties in biomechanical properties. 

\begin{table}
    \centering
    \caption{NRMSE of the voxelized flow domain using tetrahedral elements at different time instances for the descending aorta with aneurysm flow example (see Fig.10 and Fig.12).
    }
    \begin{tabular*}{\textwidth}
    {@{\extracolsep{\fill}} ccccccccc }
        \toprule
        $NRMSE$ & $t=0.1 s$ & $t=0.2 s$ & $t=0.3 s$ & $t=0.4 s$ & $t=0.5 s$ & $t=0.6 s$ & $t=0.7 s$ & $t=0.8 s$\\
        \midrule
        $u_x$ & 1.53 & 2.38 & 2.03 & 1.94 & 3.91 & 1.88 & 1.84 & 1.74  \\
        $u_y$ & 2.33 & 3.38 & 2.43 & 2.34 & 2.94 & 2.38 & 3.42 & 3.13  \\
        $u_z$ & 3.13 & 2.48 & 2.26 & 3.24 & 3.44 & 2.82 & 3.31 & 2.28 \\
        \bottomrule
    \end{tabular*}
    \label{tab:coronary-norms4}
\end{table}

\begin{figure}[ht]
    \centering
    \includegraphics[width=420pt,height=480pt]{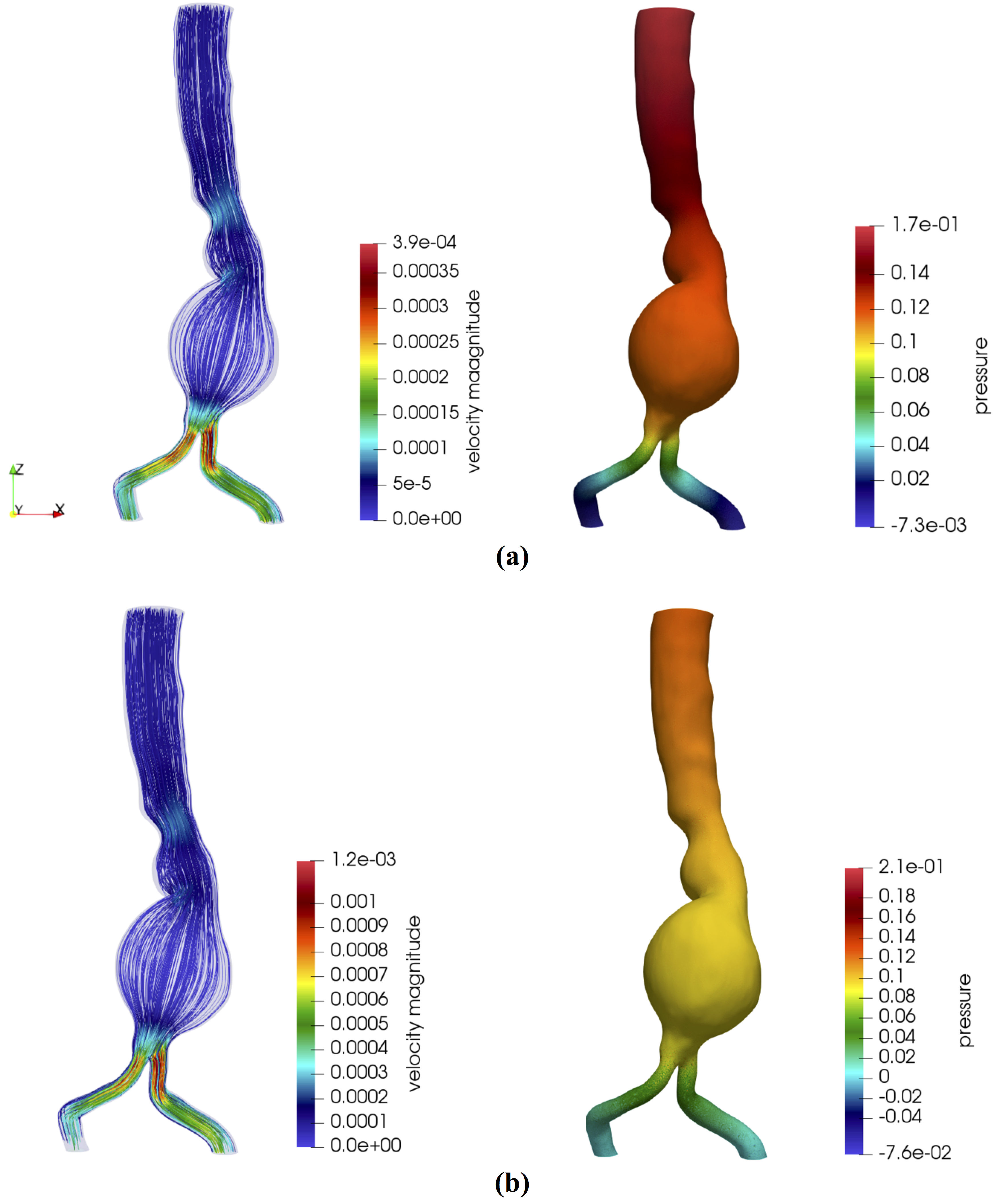}
    \caption{
        Streamlines (left) and 
         pressure (right) at
        \textbf{(a)}
        $t = 1.92$ and
        \textbf{(b)}
         $t = 2.32$ sec for the flow in the descending aorta with aneurysm example.
    }
    \label{fig:my_label}
\end{figure}

Figure 12 shows the pressure contours and the velocity streamlines at time $t_{1} = 1.92 $ sec and at $t_{2} = 2.32 $ sec. The results plotted are the numerical results computed on the voxelized mesh and then projected on the body-fitted tetrahedral mesh.

\subsection{Blood flow in the ascending aorta}
In the final example, we simulate blood flow in the ascending aorta (the brachiocephalic, left common carotid, and left subclavian arteries are included in the model) shown in Fig. 13a. 

\begin{figure}[ht]
    \centering
    \includegraphics[width=420pt,height=310pt]{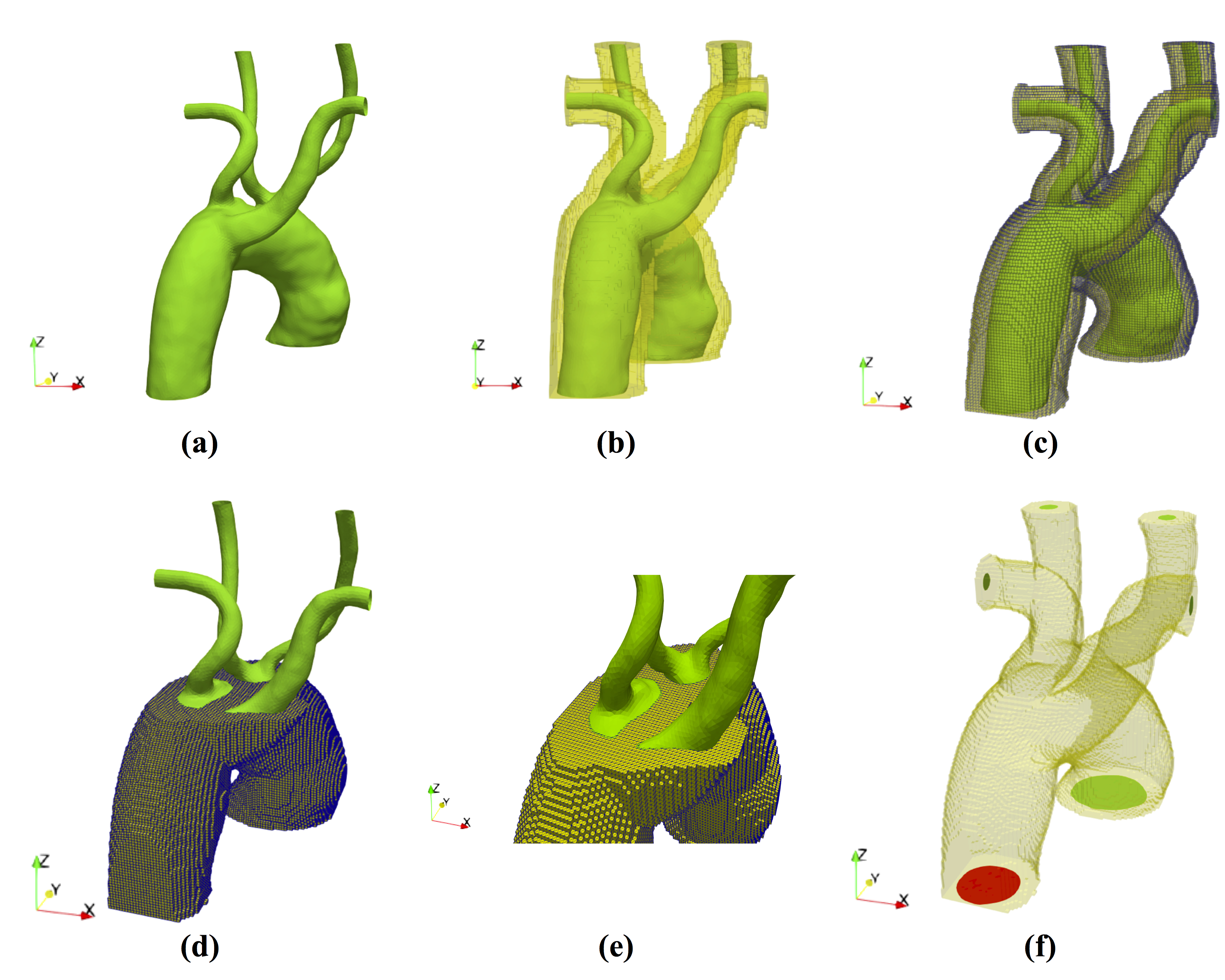}
    \caption{
        Blood flow in the ascending aorta example
        \textbf{(a)}
        immersed boundary given as STL file 
        \textbf{(b)}
        voxelized flow domain (yellow color)
        \textbf{(c)}
        immersed boundary along with the cross-section of the tetrahedral mesh
        \textbf{(d)-(e)}
        immersed boundary along with the cross-section of the tetrahedral mesh and 
        \textbf{(f)}
        flow domain for the descending aorta with aneurysm flow example, with 
        inlet (circle in red color), 
        outlets (circle in green color) and
        walls (yellow color) boundaries.
    }
    \label{fig:my_label}
\end{figure}

The voxelized flow domain (we use linear \(\left( P_{1}/P_{1} \right)\) tetrahedral elements) envelopes the immersed boundary. The immersed boundary consists of 15,938 nodes and 31,684 triangular surface elements (facets). We compute the mesh resolution $h$ of the voxelized flow domain based on the average area $\Delta S_{mean}$ (and consequently on the average length) of the surface elements (facets) of the immersed boundary. The mean area of the facets is $\Delta S_{mean} = 5.6 \times 10^{-7}$ m$^{2}$ and, therefore, the mesh resolution of the voxelized flow domain is $h = \sqrt[]{\Delta S_{mean}} = 0.00075$ m. We use six rows of elements outside the immersed boundary. We discretize the flow domain with 3,162,114 ($P_{1}/P_{1}$) elements and 558,401 nodes. This mesh ensures a grid-independent numerical solution. The voxelized mesh was generated in less than $30$ sec using a laptop computer with an i7 quad-core processor and 16 GB internal memory.

We simulate four cardiac cycles, and we set the total time of the simulation to $T = 3.44$ sec and the time step to $dt = 10^{-4}$ sec. We apply a time-dependent pressure profile at the inlet (circle in green color in Fig.13f); in the five outlets (circles in red color in Fig.13f), we apply zero pressure; and at the remaining boundaries (yellow color in Fig.13f) we apply no-slip boundary conditions. We use the Newtonian model for blood flow, with dynamic viscosity $\mu = 0.00345 Pa \cdot s$ and density of $\rho = 1,056 \frac{kg}{m^{3}}$. To demonstrate the accuracy of our IB method, we compare the results obtained using the current method---after projecting the numerical solution on a body-fitted mesh---with those for a body-fitted mesh with tetrahedral \(\left( P_{1}/P_{1} \right)\) elements (92,786 nodes and 504,628 elements). The comparison applied at different time instances (every $0.1$ s interval) of the cardiac cycle indicated a normalized root mean square error (NRMSE) of less than $5 \%$ for the three velocity components.

We visualize the results for the voxelized immersed boundary method simulation by projecting the numerical solution computed on the voxelized mesh to a body-fitted mesh used to highlight the accuracy of the proposed method. Additionally, we have conducted a qualitative comparison of the predicted streamlines and pressure contours at the time instances of $t = 1.92$ sec and $t = 2.32$ sec (Fig.14). There are no visually distinguishable differences between the results obtained using our IB method and approach using body-fitted mesh.

\begin{figure}[ht]
    \centering
    \includegraphics[width=440pt,height=360pt]{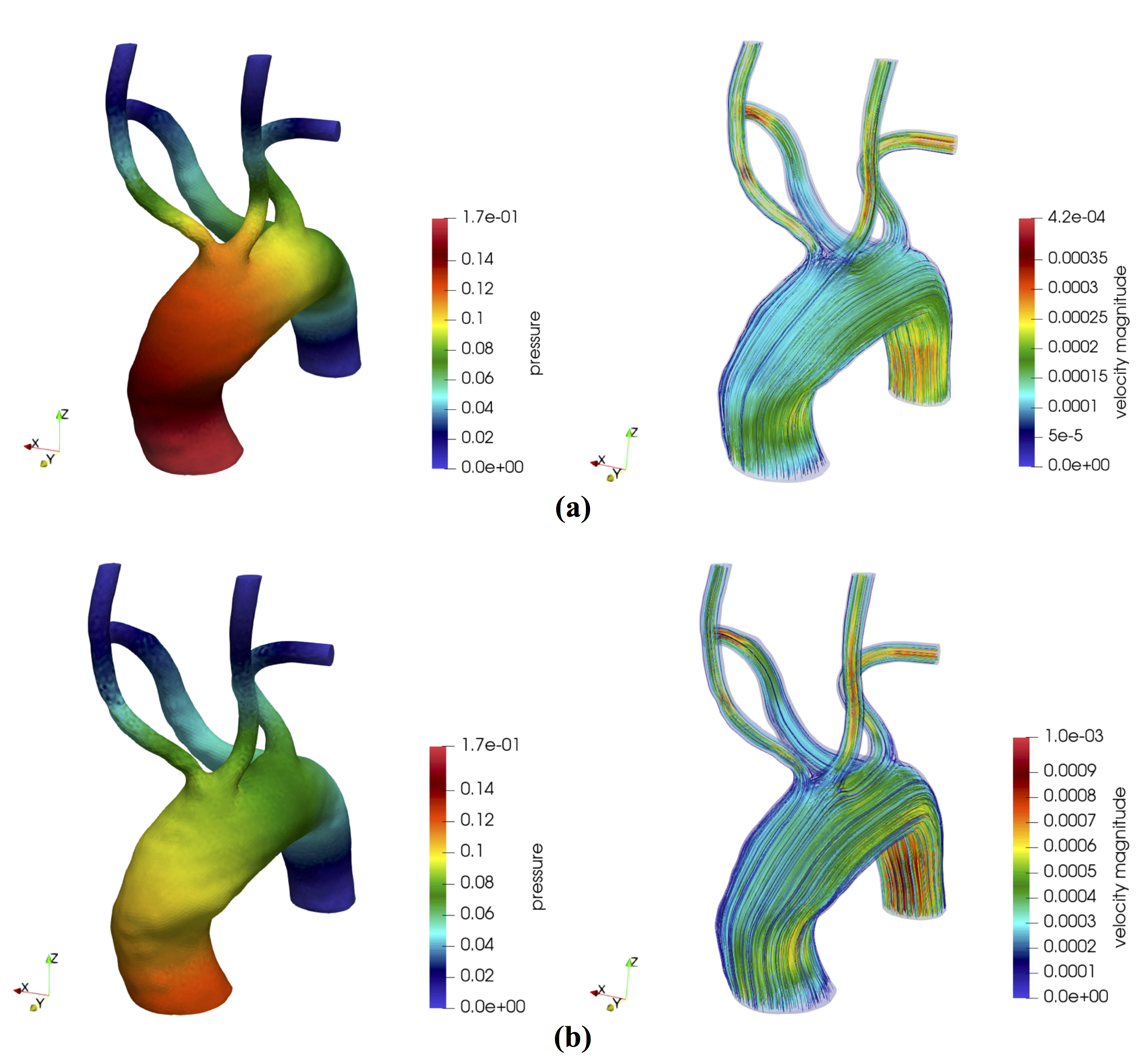}
    \caption{
        Pressure (left) and 
        streamlines (right) at
        \textbf{(a)}
        $t = 1.92$ and
        \textbf{(b)}
         $t = 2.32$ sec for the blood flow in the ascending aorta example.
    }
    \label{fig:my_label}
\end{figure}

\section{Conclusions}

In this contribution, we presented a voxelized immersed boundary method for internal
flows with application to blood flow simulations. The proposed scheme combines the finite element (FE) method to numerically solve the Navier--Stokes equations and the boundary condition enforced immersed boundary (BCE-IB) method to account for complex geometries. In this study, we focused on mesh generation and did not consider image analysis (such as segmentation and geometry reconstruction) of the blood vessels used in the simulations. The surface of blood vessels can be automatically generated from computed tomography angiography (CTA) using algorithms available in existing image processing tools such as 3D Slicer.

The proposed method increases the computational efficiency of the IB method for internal flows by decreasing the time to mesh generation and by reducing the number of unused mesh points through a sophisticated and efficient method for voxelized mesh generation in the enveloping of the immersed boundary. This method facilitated the generation of a computational mesh consisting of over three and a half million elements (computation of flow in an aortic bifurcation---see Fig. 10) in less than 30 seconds on an off-the-shelf laptop with a quad-core i7 processor and 16 GB of internal memory. These promising results make it possible to regard our flow solver as compatible with the time constraints of clinical workflows.

The proposed method utilizes an efficient and accurate finite element solver based on the incremental pressure correction scheme (IPCS). The accuracy of the proposed scheme was successfully verified through comparison with an analytical solution for the Poiseuille flow in a cylinder with a circular cross-section. We further validated the accuracy of the proposed method by solving a flow case where experimental data are available. The applicability and efficiency of the proposed method was
demonstrated through the solution of flow examples with complex geometry. We solved several problems related to the IB and FE methods. In the former, for internal flows, we use a flow domain in which most of the elements are located inside the immersed boundary, while for the latter, we use a high-quality mesh.

\clearpage
\section*{Acknowledgments}

A. Wittek and K. Miller acknowledge the support by the Australian Government through the
Australian Research Council's Discovery Projects funding scheme (project
DP160100714). The views expressed herein are those of the authors and
are not necessarily those of the Australian Research Council.

\bibliography{References}

\end{document}